\documentclass{article}
\usepackage{jheppub}
\usepackage{amsmath, amsfonts, amssymb, array}
\usepackage{xcolor}
\usepackage{mathrsfs}
\usepackage{slashed}
\newcolumntype{C}[1]{>{\centering\let\newline\\\arraybackslash\hspace{0pt}}m{#1}}
\makeatletter
\g@addto@macro\bfseries{\boldmath}
\makeatother
\usepackage[english]{babel}
\usepackage[autostyle]{csquotes}
\usepackage{float}
\usepackage{cancel}

\newcommand{\be}{\begin{equation}}
\newcommand{\ee}{\end{equation}}
\newcommand{\bea}{\begin{eqnarray}}
\newcommand{\eea}{\end{eqnarray}}
\newcommand{\nn}{\nonumber}

\usepackage{hyperref}
\newcommand{\ft}[2]{{\textstyle\frac{#1}{#2}}}
\def\rmi{{\rm i}}

\def\a{\alpha}
\def\b{\beta}

\def\m{\mu}
\def\n{\nu}
\def\r{\rho}
\def\s{\sigma}

\def\nn{\nonumber}

\def\nn{\nonumber}

\newcommand{\ben}[1]{\begin{eqnarray}\label{#1} }
\newcommand{\een}{\end{eqnarray}}

\newcommand{\DD}{{\cal D}}

\newcommand{\NN}{{\cal N}}

\newcommand{\thc}{\text{h.c.}}
\newcommand{\Dslash}{\not{\hbox{\kern-4pt $D$}}}
\newcommand{\pslash}{\not{\hbox{\kern-4pt $\partial$}}}
\newcommand{\Dcslash}{\not{\hbox{\kern-4pt $\DD$}}}

\usepackage{graphicx}
\usepackage{amssymb}
\usepackage{amsthm}
\usepackage{amsmath}
\usepackage{slashed}
\usepackage{color}
\usepackage{bbold}
\usepackage[normalem]{ulem}

\numberwithin{equation}{section}

\newcommand {\cD}{{\cal D}}

\newcommand {\cL}{{\cal L}}
\newcommand {\cM}{{\cal M}}
\newcommand {\cN}{{\cal N}}

\newcommand {\cV}{{\cal V}}

\def\a{\alpha}
\def\b{\beta}

\def\m{\mu}
\def\n{\nu}

\def\r{\rho}
\def\s{\sigma}

\makeatletter
\g@addto@macro\bfseries{\boldmath}
\makeatother

\title{Higher derivative invariants in four dimensional $\NN=3$ Poincar{\'e} Supergravity}

\author{Subramanya Hegde$^{1,2,3}$,}
\author{Madhu Mishra$^{4}$,}
\author{Debangshu Mukherjee$^{4,5}$}
\author{and Bindusar Sahoo$^{4}$}

\affiliation{$^1$ The Institute of Mathematical Sciences, Taramani, Chennai - 600113, India \\
	$^2$ Harish-Chandra Research Institute, Jhunsi, Allahabad - 211019, India \\
	$^3$ Homi Bhabha National Institute, Anushakti Nagar, Mumbai - 400085, India\\
	$^4$ Indian Institute of Science Education and Research,
	Vithura, Thiruvananthapuram - 695551, India\\
	$^5$ Department of Physics, Indian Institute of Technology Kanpur, Kanpur - 208016, India}

\emailAdd{subbuh@imsc.res.in}
\emailAdd{madhu50315@iisertvm.ac.in}
\emailAdd{debangshu@iitk.ac.in}
\emailAdd{bsahoo@iisertvm.ac.in}

\abstract{In this paper we use the superconformal approach to derive the higher derivative action for $\NN=3$ Poincar{\'e} supergravity in four space-time dimensions. We first study the coupling of $\NN=3$ vector multiplets to conformal supergravity. Thereafter we combine it with the pure $\NN=3$ conformal supergravity action and use a minimum of three vector multiplets as compensators to arrive at Poincar{\'e} supergravity with higher derivative corrections. We give a general prescription on how to eliminate the auxiliary fields in an iterative manner and obtain the supergravity action order by order in derivatives. We also show that the truncation of the action at fourth order in derivatives is a consistent truncation.}

\begin{document}
	\allowdisplaybreaks
	\maketitle
	\section{Introduction}
		Supersymmetry offers a window into non-perturbative physics. While $\mathcal{N}=1,2,4$ supersymmetric theories in four dimensions have been studied extensively, it has been known that perturbatively $\mathcal{N}=3$ theories with rigid supersymmetry are disguised versions of theories with $\mathcal{N}=4$ supersymmetry \cite{weinberg2005quantum}. To see this, one can simply look at the massless matter representations of $\mathcal{N}=3$ supersymmetry where, once we add the CPT conjugate representation to ensure that the \emph{full} multiplet is CPT invariant, the field content matches with that of $\mathcal{N}=4$ theory. This can be used to see that perturbatively, there can not be theories that exhibit genuine rigid $\mathcal{N}=3$ supersymmetry.  However, in recent years several non-perturbative constructions have been found which evade this, where the theories have no weak coupling limit \cite{Aharony:2015oyb,Garcia-Etxebarria:2015wns,Nishinaka:2016hbw,Aharony:2016kai,Imamura:2016abe,Agarwal:2016rvx,Garcia-Etxebarria:2016erx,Lemos:2016xke,Bourton:2018jwb}. For supergravity theories, the situation is different from the outset. If we consider the massless representation of $\mathcal{N}=3$ supersymmetry with maximal helicity $2$, which corresponds to supergravity theories, the field content no longer matches that of $\mathcal{N}=4$ supergravity theories. Thus, there exist supergravity theories which exhibit genuine $\mathcal{N}=3$ supersymmetry. Constructing these supergravity theories is important in the context of classifying supergravity theories in four dimensions. Further, in the light of gauge/gravity duality, $\mathcal{N}=3$ $AdS_4$ solutions and the associated field representations were studied in \cite{Fre:1999gok,Termonia:1999cs,Castellani:1983tc} and have seen renewed interest \cite{Ferrara:1998zt,Cheon:2011th,Karndumri:2016tpf,Karndumri:2016miq,Karndumri:2021jln,Fre:2022ncd}. This provides the motivation to construct higher derivative invariants in $\NN=3$ supergravity as they are known to provide important insights for holography \cite{Bobev:2020egg,Bobev:2021oku,Genolini:2021urf}.
	
	The study of supergravity theories with higher derivative corrections is greatly facilitated by the superconformal approach. The superconformal approach relies on a larger group of symmetries: the superconformal symmetry, which distributes the physical degrees of freedom in shorter multiplets and therefore it becomes technically tractable to construct such theories. In this framework, one first studies a theory of conformal supergravity coupled to additional matter multiplets and then uses some of these matter multiplets to gauge fix the extra symmetries to arrive at a supergravity theory with only super-Poncar{\'e} symmetry. 
	
	$\mathcal{N}=1,2,4$ supergravity theories have been well studied in the literature using the superconformal formalism. However, $\mathcal{N}=3$ supergravity theories are relatively less studied Previous works looked at supergravity solutions and field content of $\mathcal{N}=3$ theories that arose in compactifications of certain 11D supergravity theories on so-called tri-Sasakian manifolds \cite{Termonia:1999cs,Cassani:2011fu,Karndumri:2016tpf}. Further, at two derivative level, matter couplings to Poincar\'e supergravity was presented in \cite{Castellani:1985ka}. Superspace approaches have also been developed \cite{Galperin:1986id,Howe:1980sy,Brink:1978sz}. However, to construct general higher derivative matter couplings, it is instructive to look at $\mathcal{N}=3$ supergravity using the superconformal approach. In this approach,  a crucial ingredient is the $\mathcal{N}=3$ Weyl multiplet whose components were conjectured long back in \cite{Fradkin:1985am}. However, it was only recently the explicit transformation laws of the component fields under the $\mathcal{N}=3$ superconformal symmetry were written down \cite{vanMuiden:2017qsh,Hegde:2018mxv}. Subsequently, action for the $\mathcal{N}=3$ conformal supergravity theory was constructed in \cite{Hegde:2021rte}. This was done by first constructing a ``density formula''\footnote{Density formula basically gives an invariant action in terms of an abstract multiplet. Using this formula one can construct an invariant action for any known multiplet by simply embedding the known multiplet in the abstract multiplet.} using the ``covariant superform action principle"\footnote{This principle is the same as the ``ectoplasm principle'' \cite{Gates:1997ag,Gates:1997kr} and ``rheonomy principle'' \cite{DAuria:1982mkx,Castellani:1991eu} and  was used recently also in \cite{Butter:2016mtk,Butter:2019edc,Hegde:2019ioy}.}. Following this, the Weyl multiplet was embedded in the density formula to derive an explicit form of the $\mathcal{N}=3$ conformal supergravity action. The work of \cite{Hegde:2021rte} also found a consistent truncation of the $\mathcal{N}=4$ Weyl multiplet to the $\mathcal{N}=3$ Weyl multiplet and used it to check that the $\mathcal{N}=3$ conformal supergravity action indeed arise from the truncation of the $\mathcal{N}=4$ conformal supergravity action derived in \cite{Butter:2016mtk,Butter:2019edc}.
	
	The aim of this paper is to look at pure $\mathcal{N}=3$ Poincar{\'e} supergravity using the superconformal approach and construct higher derivative invariants. To achieve this, we need a compensating multiplet to gauge fix additional symmetries in the superconformal theory. The on-shell $\cN=3$ vector multiplet field content has been known \cite{Castellani:1985ka,Karndumri:2016miq} and there is a harmonic superspace construction to render the multiplet off-shell \cite{Galperin:1984bu,Galperin:1985uw,galperin2001harmonic}. We begin our work by finding the transformation law of the on-shell $\mathcal{N}=3$ vector multiplet coupled to conformal supergravity. This is done by performing a supersymmetry reduction of the $\mathcal{N}=4$ vector multiplet coupled to conformal supergravity. As we will see later, the components of the on-shell $\mathcal{N}=3$ vector multiplet are not truncations but rather mere re-arrangements of the components of the $\mathcal{N}=4$ vector multiplet which is obvious as we have discussed earlier, due to CPT invariance. However the key difference arises in the way it couples to conformal supergravity since the $\mathcal{N}=3$ Weyl multiplet is fundamentally different from the $\mathcal{N}=4$ Weyl multiplet. We follow the same truncation scheme as followed in \cite{Hegde:2021rte}, where we set one of the supersymmetries to zero while the $SU(1,1)$ coset scalars $\phi_{\alpha}$ that are present in the $\mathcal{N}=4$ theory are set to the constant values of $\phi_1=1$ and $\phi_2=0$. This provides a map between the various components of the vector multiplets between the $\mathcal{N}=3$ and $\mathcal{N}=4$ theories. Thereafter, we follow \cite{Hegde:2021rte} and embed the vector multiplet in the density formula to obtain an action for the $\mathcal{N}=3$ vector multiplet coupled to conformal supergravity. However, this procedure will give only a partial action due to the on-shell nature of the vector multiplet. Such a feature is akin to vector multiplet coupled to $\mathcal{N}=4$ conformal supergravity, where again the embedding of vector multiplet in the $\mathcal{N}=4$ density formula leads to mismatches in certain kinetic terms \cite{Butter:2019edc}. The ``density formula'' derived from the ``superform action principle'' relies on the closure of the supersymmetry algebra. Hence as long as it is being used to derive an action for a multiplet on which the algebra closes off-shell, such as the Weyl multiplet, then it produces the correct action. But if one wants to apply it on an on-shell multiplet where the algebra closes modulo equations of motion, such as $\mathcal{N}=3,4$ vector multiplets, then the action one obtains may be blind to terms that vanish when one uses the equations of motion. Hence in order to find the full action, one needs to know the equations of motion of the $\mathcal{N}=3$ vector multiplet. This is obtained by using our truncation scheme on the equations of motion of the $\mathcal{N}=4$ vector multiplet coupled to conformal supergravity derived in \cite{deRoo:1984zyh}. Further, one adds terms proportional to the equations of motion to the partial action obtained using the density formula and demands that the same equations of motion arise from a variational principle of the modified action. In this way one may obtain the full action of $\mathcal{N}=3$ vector multiplet coupled to conformal supergravity. This action also straightforwardly generalizes to an action for $n_{v}$ vector multiplets coupled to conformal supergravity. This generalization is done carefully so that the action for three vector multiplets come with the wrong sign of the kinetic terms so that they can be used as compensators to go from conformal supergravity to Poincar{\'e} supergravity by choosing appropriate gauge fixing conditions for the extra symmetries (dilatation, special conformal transformation, $SU(3)\times U(1)$ R-symmetry and conformal-supersymmetry or special (S)-supersymmetry). 

	To summarize, we start Sec. \ref{sec:weyl} by briefly reviewing $\cN=3$ conformal supergravity, where we discuss the field content of the Weyl multiplet in $\mathcal{N}=3$ conformal supergravity, the $\cN=3$ density formula and the truncation of the $\mathcal{N}=4$ Weyl multiplet to its $\mathcal{N}=3$ counterpart based on previous work \cite{Hegde:2021rte}. In Sec. \ref{sec:vector}, we introduce the $\mathcal{N}=3$ vector multiplet along with its transformation rule under ordinary ($Q$) and special ($S$)-supersymmetry.  We also perform the reduction of the equations of motion of the $\mathcal{N}=4$ vector multiplet coupled to conformal supergravity to their $\mathcal{N}=3$ counterparts using the truncation scheme discussed earlier. In the next section, Sec. \ref{sec:embedding}, we discuss the embedding of the vector multiplet in the $\mathcal{N}=3$ density formula and derive the Lagrangian for Maxwell action coupled to $\mathcal{N}=3$ conformal supergravity. We will then give a straightforward generalization of this action for $n_v$ vector multiplets coupled to conformal supergravity in such a way that the action for three of the vector multiplets come with the wrong sign of the kinetic terms and the remaining come with the right sign. Since in this paper, we will be interested in pure Poincar{\'e} supergravity, we will restrict ourselves to $n_{v}=3$ with all of them coming with the wrong sign of the kinetic terms. In Sec. \ref{sec:gaugefixing}, we will demonstrate how we can use the compensating vector multiplets for explicit breaking of the extra symmetries present in the superconformal theory and arrive at supergravity with $\mathcal{N}=3$ super-Poincar{\'e} symmetries.  In Sec. \ref{HDD}, we will also show how we can eliminate the auxiliary fields in an iterative fashion and obtain the action as an expansion in the number of derivatives. We will also demonstrate how the truncation of the action to the fourth order in the derivative is a consistent truncation. Finally, we will end with some conclusions and discussions.

	\section{$\cN=3$ conformal supergravity}\label{sec:weyl}
	
	A conformal supergravity theory with minimal/extended $(\cN\geq1)$ supersymmetry in four dimensions is a gravity theory that is obtained from the gauge theory based on the superconformal algebra $su(2,2|\cN)$. The gauge theory transforms into a theory of gravity upon imposing an appropriate set of curvature constraints. The Weyl multiplet in conformal supergravity is the multiplet that contains the gauge fields of the superconformal algebra. For extended supersymmetry ($\cN>1$), one also requires an extra set of auxiliary fields for the off-shell closure of the algebra. The algebra is also modified where the structure constants become field dependent, and this leads to a soft superconformal algebra. For more details, please refer to \cite{Freedman:2012zz}. The  $\cN=3$ Weyl multiplet in four dimensions was constructed in \cite{vanMuiden:2017qsh, Hegde:2018mxv}, and we will discuss its structure in the following subsection.
	
	\subsection{$\cN=3$ Weyl multiplet}\label{sec-density}
	The $\cN=3$ Weyl multiplet is a $64+64$ (bosonic+fermionic) multiplet whose components are as tabulated in Table-\ref{Table-Weyl}.
	\begin{table}[ht]
		\caption{Field content of the $\cN = 3$ Weyl multiplet}\label{Table-Weyl}
		\begin{center}
			\begin{tabular}{ | C{2cm}|C{2cm}|C{3cm}|C{2cm}|C{2cm}| }
				\hline
				Field & SU(3) Irreps & Restrictions &Weyl weight (w) & Chiral weight (c) \\ \hline
				$e_{\mu}{}^{a}$ & $\bf{1}$ & Vielbein & -1 & 0 \\ \hline
				$V_{\mu}{}^{i}{}_{j}$ & $\bf{8}$ & $(V_{\mu}{}^{i}{}_{j})^{*}\equiv V_{\mu}{}_{i}{}^{j}=-V_{\mu}{}^{j}{}_{i}$ SU(3)$_R$ gauge field &0 & 0  \\ \hline
				$A_{\mu}$ & $\bf{1}$ & U(1)$_R$ gauge field &0 & 0  \\ \hline
				$b_{\mu}$ & $\bf{1}$ & dilatation gauge field &0 & 0  \\ \hline
				$T^{i}_{ab}$ & $\bf{3}$ & Self-dual i.e $T^{i}_{ab}=\frac{1}{2}\varepsilon_{abcd}T^{i}{}^{cd}$ &1 & 1  \\ \hline
				$E_{i}$ &$\bf{\bar{3}}$ & Complex & 1&-1\\ \hline
				$D^{i}{}_{j}$ & $\bf{8}$ & $(D^{i}{}_{j})^{*}\equiv D_{i}{}^{j}=D^{j}{}_{i}$ &2 & 0  \\ \hline
				$\psi_{\mu}{}^{i}$ & $\bf{3}$ & $\gamma_{5}\psi_{\mu}{}^{i}=\psi_{\mu}{}^{i}$&-1/2 & -1/2  \\ \hline
				$\chi_{ij}$ & $\bf{\bar{6}}$ & $\gamma_{5}\chi_{ij}=\chi_{ij}$ &3/2 & -1/2  \\ \hline
				$\zeta^{i}$ & $\bf{3}$ & $\gamma_{5}\zeta^{i}=\zeta^{i}$ & 3/2 &-1/2 \\ \hline
				$\Lambda_{L}$ & $\bf{1}$ & $\gamma_{5}\Lambda_{L}=\Lambda_{L}$ &1/2 &-3/2 \\ \hline
			\end{tabular}
		\end{center}
	\end{table}
	There are two types of supersymmetry in conformal supergravity: $Q$ (or ordinary) supersymmetry and $S$ (or special) supersymmetry. The $Q$ and $S$-supersymmetry transformations of the components of the Weyl multiplet are given as:
	\begin{align}\label{N3susy}
	\delta e_{\mu}^{a}&= \bar{\epsilon}_{i}\gamma^{a}\psi_{\mu}^{i}+\thc \nonumber \\
	\delta \psi_{\mu}^{i}&=2\mathcal{D}_{\mu}\epsilon^{i}-\frac{1}{8}\varepsilon^{ijk}\gamma\cdot T_{j}\gamma_{\mu}\epsilon_{k}-\varepsilon^{ijk}\bar{\epsilon}_{j}\psi_{\mu k}\Lambda_{L}-\gamma_{\mu}\eta^{i} \nonumber \\
	\delta V_\mu{}^i{}_j &=\bar{\epsilon}^i\phi_{\mu j}- \frac{1}{48}\bar{\epsilon}^i\gamma_\mu\zeta_j+ \frac{1}{16}\varepsilon_{jkl}\bar{\epsilon}^k\gamma_\mu\chi^{il}- \frac{1}{16}\bar{\epsilon}^i\gamma\cdot T_j \gamma_\mu\Lambda_R- \frac{1}{16}\bar{\epsilon}^i\gamma_\mu \Lambda_R E_j +\frac{1}{8}\varepsilon_{klj}E^i\bar{\epsilon}^k\psi_\mu^l \nonumber \\
	&\quad+ \frac{1}{4}\bar{\epsilon}^i\gamma^a\psi_{\mu j}\bar{\Lambda}_L\gamma_a\Lambda_R-\bar{\psi}_\mu^i\eta_j-\thc-\text{trace} \nonumber \\
	\delta A_\mu &=\frac{i}{6}\bar{\epsilon}^i\phi_{\mu i}+ \frac{i}{36}\bar{\epsilon}^i\gamma_\mu\zeta_i+ \frac{i}{12}\varepsilon_{klp}E^p\bar{\epsilon}^k\psi_{\mu}^l+ \frac{i}{12}\bar{\epsilon}^i\gamma\cdot T_i\gamma_\mu\Lambda_R+\frac{i}{12}\bar{\epsilon}^i\gamma_\mu\Lambda_RE_i-\frac{i}{3}\bar{\epsilon}^i\gamma^a\psi_{\mu i}\bar{\Lambda}_L\gamma_a\Lambda_R\nonumber\\
	&\quad-\frac{i}{6}\bar{\psi}_\mu^i\eta_i+\thc \nonumber \\
	\delta b_\mu &= \frac{1}{2}(\bar{\epsilon}^i\phi_{\mu i}-\bar{\psi}_\mu^i\eta_i)+\thc\nonumber \\
	\delta \Lambda_L&=-\frac{1}{4}E_i\epsilon^i+\frac{1}{4}\gamma\cdot T_i\epsilon^i\nonumber \\
	\delta E_i &=-4 \bar{\epsilon}_i\slashed{D}\Lambda_L-\frac{1}{2}\varepsilon_{ijk}\bar{\epsilon}^j\zeta^k+\frac{1}{2}\bar{\epsilon}^j\chi_{ij}-\frac{1}{2}\varepsilon_{ijk}E^k\bar{\epsilon}^j\Lambda_L-4\bar{\Lambda}_L\Lambda_L\bar{\epsilon}_i\Lambda_R- 4\bar{\eta}_i\Lambda_L\nonumber \\
	\delta T^i_{ab} &= -\bar{\epsilon}^i\slashed{D}\gamma_{ab}\Lambda_R-4\varepsilon^{ijk}\bar{\epsilon}_jR_{ab}(Q)_k+\frac{1}{8}\bar{\epsilon}_j\gamma_{ab}\chi^{ij}+\frac{1}{24}\varepsilon^{ijk}\bar{\epsilon}_j\gamma_{ab}\zeta_k-\frac{1}{8}\varepsilon^{ijk}E_j\bar{\epsilon}_k\gamma_{ab}\Lambda_R\nonumber \\
	&\quad+\bar{\eta}^i\gamma_{ab}\Lambda_R\nonumber \\
	\delta \chi_{ij}&=2\slashed{D}E_{(i}\epsilon_{j)}-8\varepsilon_{kl(i}\gamma\cdot R(V)^l{}_{j)}\epsilon^k-2\gamma\cdot\slashed{D}T_{(i}\epsilon_{j)}+\frac{1}{3}\varepsilon_{kl(i}D^l{}_{j)}\epsilon^k\nonumber \\&\quad+\frac{1}{4}\varepsilon_{kl(i}E^k\gamma\cdot T_{j)}\epsilon^l-\frac{1}{3}\bar{\Lambda
	}_L\gamma_a\epsilon_{(i}\gamma^a\zeta_{j)}+\frac{1}{4}\varepsilon_{lm(i}E_{j)}E^m\epsilon^l-\bar{\Lambda}_L\gamma^a\Lambda_R\gamma_aE_{(i}\epsilon_{j)}\nonumber\\
	&\quad-\bar{\Lambda}_L\gamma\cdot T_{(i}\gamma^a\Lambda_R\gamma_a\epsilon_{j)}+ 2\gamma\cdot T_{(i}\eta_{j)}+ 2E_{(i}\eta_{j)}\nonumber \\
	\delta \zeta^i &=- 3\varepsilon^{ijk}\slashed{D}E_j\epsilon_k +\varepsilon^{ijk}\gamma\cdot\slashed{D}T_k\epsilon_j-4\gamma\cdot R(V)^i{}_j\epsilon^j-16i\gamma\cdot R(A)\epsilon^i-\frac{1}{2}D^i{}_j\epsilon^j-\frac{3}{8}E^i\gamma\cdot T_j\epsilon^j\nonumber\\
	&\quad+\frac{3}{8}E^j\gamma\cdot T_j\epsilon^i+\frac{3}{8}E^iE_j\epsilon^j+\frac{1}{8}E^jE_j\epsilon^i\nonumber\\
	&\quad- 4 \bar{\Lambda}_L\slashed{D}\Lambda_{R}\epsilon^i- 4 \bar{\Lambda}_R\slashed{D}\Lambda_L\epsilon^i- 3\bar{\Lambda}_R\slashed{D}\gamma_{ab}\Lambda_L\gamma^{ab}\epsilon^i-3\bar{\Lambda}_L\gamma_{ab}\slashed{D}\Lambda_R\gamma^{ab}\epsilon^i\nonumber\\
	&\quad+\frac{1}{2}\varepsilon^{ijk}\bar{\Lambda}_L\gamma^a\epsilon_j\gamma_a\zeta_k-6\bar{\Lambda}_L\Lambda_L\bar{\Lambda}_R\Lambda_R\epsilon^i+\varepsilon^{ijk}\gamma\cdot T_j\eta_k-3\varepsilon^{ijk}E_j\eta_k\nonumber\\
	\delta D^i{}_j&=-3\bar{\epsilon}^i\slashed{D}\zeta_j-3\varepsilon_{jkl}\bar{\epsilon}^k\slashed{D}\chi^{il}+\frac{1}{4}\varepsilon_{jkl}\bar{\epsilon}^i\zeta^k E^l+\frac{1}{2}\varepsilon_{jkl}\bar{\epsilon}^k\zeta^l E^i+\frac{3}{4}\bar{\epsilon}^i\chi_{jk}E^k+ 3\bar{\epsilon}^i\gamma\cdot T_j\overset{\leftrightarrow}{\slashed{D}}\Lambda_R\nonumber\\
	&\quad-3\bar{\epsilon}^i\slashed{D}\Lambda_RE_j-3\bar{\epsilon}^i\slashed{D}E_j\Lambda_R+ \frac{3}{4}\varepsilon_{jkl}E^l\bar{\epsilon}^k\Lambda_LE^i+ {3\varepsilon_{jkl}T^i\cdot T^l\bar{\epsilon}^k\Lambda_{L}}-2\bar{\epsilon}^i\Lambda_L\bar{\Lambda}_R\zeta_j\nonumber\\
	&\quad-3\bar{\epsilon}^i\Lambda_L\bar{\Lambda}_R\Lambda_RE_j+3\bar{\epsilon}^i\gamma\cdot T_j\Lambda_L\bar{\Lambda}_R\Lambda_R+\thc-\text{trace}
	\end{align}
	where, $\mathcal{D}_{\mu}\epsilon^{i}$ is defined as:
	\begin{align}\label{Depsilon}
	\mathcal{D}_\mu\epsilon^i=\partial_\mu\epsilon^i-\frac{1}{4}\gamma\cdot\omega_\mu\epsilon^i+\frac{1}{2}(b_\mu+iA_\mu)\epsilon^i-V_\mu{}^i{}_j\epsilon^j 
	\end{align}
	The covariant derivative $D_a$ that appears above is fully supercovariant w.r.t all the gauge transformations (bosonic as well as fermionic). The covariant derivative $\mathcal{D}_{a}$ is often used to denote covariantization only w.r.t the ``standard gauge transformations'' which are dilatation (D), local Lorentz transformation (M), $SU(3)$ R-symmetry (V) and $U(1)$ R-symmetry (A). Apart from the above mentioned standard gauge transformations, which are bosonic, there is another non-standard bosonic gauge transformation which is the special conformal transformation (K). The complete $Q$ and $S$-supersymmetry transformations of the dependent gauge fields corresponding to local Lorentz transformations ($\omega_{\mu}^{ab}$), $S$-supersymmetry ($\phi_{\mu}^{i}$) and special conformal transformation or $K$-gauge field ($f_{\mu}^{a}$), along with the constraints on the curvatures arising from the Bianchi identities, are provided in appendix \ref{trans_dep}.

	\subsection{$\cN=3$ Density formula}\label{density-formula}
	Superconformally invariant actions in conformal supergravity are obtained by exploiting the knowledge of density formulae. Density formulae in conformal supergravity are superconformally invariant actions given either in terms of an abstract multiplet or a known multiplet. For example, the well known chiral density formula in $\cN=2$ conformal supergravity is built on an $\cN=2$ chiral multiplet \cite{deRoo:1980mm}. In this subsection, we will review the density formula and transformation rules for the relevant fields of the abstract multiplet in the context of $\NN=3$ conformal supergravity based on \cite{Hegde:2021rte}. The density formula in $\cN = 3$ theory is constructed from an abstract multiplet using the superform action principle  \cite{Butter:2019edc, Hegde:2019ioy, Hegde:2021rte}; for details on the construction, see  \cite{ Hegde:2021rte}. The density formula is given as the integral of the following 4-form:

	\be
	S = \int J,
	\ee
	where the 4-form $J$ is given in terms of the composites constructed out of the components of the abstract multiplet as,
	\begin{align}\label{full-density}
	J&=\frac{1}{72}e^a e^b e^c e^d \boldsymbol{\mathcal{L}} \varepsilon_{abcd}+ \frac{1}{3}e^a e^b e^c \bar{\psi}^{k}\gamma^{d} \boldsymbol{\mathcal{N}_{k}} \varepsilon_{abcd}+\frac{1}{3}e^a e^b e^c \bar{\psi}^{k} \boldsymbol{\mathcal{M}^{d}_{k}} \varepsilon_{abcd}\nn\\
	&\quad-\frac{1}{12}e^{a}e^{b}\bar{\psi}^{l}\gamma^{c}\psi_{k}\bar{\Lambda}_{L}\gamma^{d} \boldsymbol{\rho^{k}{}_{l}} \varepsilon_{abcd}-\frac{1}{2}e^a e^b \bar{\psi}^{i}\psi^{j} \boldsymbol{H^{-}{}^{l}_{ab}} \varepsilon_{ijl}-\frac{1}{2}e^a e^b \bar{\psi}^{i}\gamma_{ab}\psi^{j}\boldsymbol{K_{ij}}\nn\\
	&\quad-\frac{1}{32}e^a e^b \bar{\psi}^{i}\psi^{j}\boldsymbol{G^{+}{}^{l}_{ab}}\varepsilon_{ijl}-\frac{1}{2}e^{b}\bar{\psi}^{i}\psi^{j}\bar{\psi}_{k}\gamma_{b}\boldsymbol{C^{kl}}\Lambda_L\varepsilon_{ijl}-\frac{1}{6}\varepsilon_{kln}e^{a}\bar{\psi}^{k}\psi^{l}\bar{\psi}^{m}\gamma_{a}\boldsymbol{\rho^{n}{}_{m}}\nn\\
	&\quad+\frac{1}{4}\bar{\psi}^{i}\psi^{j}\bar{\psi}^{k}\psi^{l}\varepsilon_{ijm}\varepsilon_{kln}\boldsymbol{C^{mn}}+\thc
	\end{align}
	The vielbein and gravitino 1-forms are denoted as $e^a$ and $\psi^{i}$, respectively. The wedge product between the forms is suppressed in the above expression. The fields represented in bold are the composites constructed out of the components of the abstract multiplet. The composite $C^{ij}$ appearing with the maximum number of gravitino 1-forms is the lowest component of the abstract multiplet, has Weyl weight $+2$, chiral weight $+2$, and is invariant under S-supersymmetry. It is in the $\bf{6}$ of SU(3) and its complex conjugate $C_{ij}$ is in the $\bf{\bar{6}}$ of SU(3). It also needs to satisfy the constraints
	\begin{align}\label{C_constraints}
	\left(\nabla_{k}C^{ij}\right)_{\textbf{15}}&=0\nn\\
	\nabla^{(i}C^{jk)}&=0
	\end{align}
	where $\nabla_{k}$ and $\nabla^k$ are the generators for the left chiral Q-supersymmetry and right chiral Q-supersymmetry, respectively. In other words, the left and right chiral Q-supersymmetry transformation of $C^{ij}$ should take the following form:
	\begin{align}\label{left_right_Cij}
	\delta^Q_L C^{ij}&=\frac{1}{2}\bar{\epsilon}^{(i}\hat{\rho}^{j)}\nn\\
	\delta^{Q}_{R}C^{ij} &=\frac{2}{3}\varepsilon^{lk(i}\bar{\epsilon}_{k}\rho^{j)}{}_{l}
	\end{align}
	The component $\rho^{i}{}_{j}$ appears in one of the cubic gravitino terms of the density formula (\ref{full-density}). As shown below, the component $\hat{\rho}^{j}$ appears in the composite $H^{-}{}^{l}_{ab}$ which appears in one of the quadratic gravitino terms:
	\begin{align}\label{3.35}
	H^{-}{}^{l}_{ab}&=\frac{1}{2}C^{lm}T_{ab m}-\frac{1}{16}\bar{\Lambda}_{L}\gamma_{ab}\hat{\rho}^{l},
	\end{align}
	\noindent
	The composite $K_{ij}$ that appears in another quadratic gravitino term is given as:
	\begin{align}\label{KG}
	K_{ij}&=\frac{1}{24}F_{ij}+\frac{1}{4}\bar{\Lambda}_{R}\Lambda_{R}C_{ij}
	\end{align}
	The term $F_{ij}$ that appears above and the composite $G^{+l}_{ab}$ that appears in another quadratic gravitino term in the density formula arises upon the application of the right chiral Q-supersymmetry transformation on $\rho^{i}{}_{j}$ as shown below:
	\begin{align}\label{right_susy_rho}
	\delta_{Q}^{R}\rho^{i}{}_{j}&=\frac{3}{4}C^{ik}E_{j}\epsilon_{k}-\frac{1}{4}\delta^{i}_{j}C^{lk}E_{k}\epsilon_{l}+\frac{3}{8}\bar{\Lambda}_{L}\hat{\rho}^{i}\epsilon_{j}-\frac{1}{8}\delta^{i}_{j}\bar{\Lambda}_{L}\hat{\rho}^{k}\epsilon_{k}\nonumber\\
	&\quad -\frac{1}{4}\varepsilon^{ikl}F_{jk}\epsilon_{l}+\frac{3}{64}\gamma\cdot G^{i}\epsilon_{j}-\frac{1}{64}\delta^{i}_{j}\gamma\cdot G^{k}\epsilon_{k}
	\end{align}
	The composite $\mathcal{N}_{k}$ and $\mathcal{M}_{ak}$ that appears in the linear gravitino terms are given as:
	\begin{align}\label{MN-composites}
	{\mathcal{N}}_{k}&=-\frac{1}{32}\gamma\cdot T^{l}\Lambda_{R}C_{kl}+\frac{1}{192}\tilde{\theta}_{k}+\frac{1}{4}\theta_{k}\nn\\
	\mathcal{M}_{ak}&=-\frac{1}{64}\gamma\cdot T_{j}\gamma_{a}\rho^{j}{}_{k}+\frac{1}{48}\Upsilon_{ak}
	\end{align}
	The components $\theta_k$, $\tilde{\theta}_{k}$ and $\Upsilon_{ak}$ appearing above arises upon taking the right-supersymmetry transformations on $K_{ij}$  and left-supersymmetry transformation on the composite $\mathcal{G}_{a}{}^{i}{}_{j}\equiv \bar{\Lambda}_{L}\gamma_{a}\rho^{i}{}_{j}-\text{h.c}$, as shown below:
	\begin{align}\label{KG-transformation}
	\delta_{Q}^{R}K_{ij}&=\bar{\epsilon}_{k}\tau^{k}_{ij}+\frac{1}{2}\bar{\epsilon}_{(i}\theta_{j)}\nonumber\\
	\delta_{Q}^{L}\mathcal{G}_{d}{}^{i}{}_{j}&=\bar{\epsilon}^{k}\alpha_{d}{}^{i}_{jk}+\frac{1}{4}\bar{\epsilon}^{k}\gamma_{d}\tilde{\alpha}^{i}_{jk}+\frac{1}{2}\varepsilon_{mjk}\varepsilon^{m}\beta_{d}^{ik}+\frac{1}{8}\varepsilon_{mjk}\varepsilon^{m}\gamma_{d}\tilde{\beta}^{ik}+\frac{3}{8}\bar{\epsilon}^{i}{\Upsilon}_{dj}\nonumber\\
	&\;\;\; +\frac{3}{32}\bar{\epsilon}^{i}\gamma_{d}\tilde{\theta}_{j}-\frac{1}{8}\delta^{i}_{j}\bar{\epsilon}^{m}{\Upsilon}_{dm}-\frac{1}{32}\delta^{i}_{j}\bar{\epsilon}^{m}\gamma_{d}\tilde{\theta}_{m}
	\end{align}
	And finally the composite $\mathcal{L}$ that appears in the $e^4$ term in the density formula is given as:		
	\begin{align}\label{comp_L}
	\mathcal{L}&=Y+\frac{1}{2}H^{-l}\cdot T_{l}-\text{h.c}
	\end{align}
	where $H^{-l}_{ab}$ is already defined in (\ref{3.35}) and the component $Y$ appears in the right-supersymmetry transformations of $\mathcal{N}_{i}$ as shown below:
	\begin{align}\label{3.40}
	\delta_{Q}^{R}\mathcal{N}_{i}&=-\frac{1}{2}W^{j}{}_{i}\epsilon_{j}-\frac{1}{6}Y\epsilon_{i}+\frac{1}{8}Z^{+}_{ab}{}^{j}{}_{i}\epsilon_{j}+\frac{1}{24}\tilde{Z}^{+}_{ab}\epsilon_{i}
	\end{align}
\noindent	
	In order to find an invariant action for a known multiplet in $\cN=3$ conformal supergravity, one needs to find the lowest component $C^{ij}$ of the abstract multiplet as a composite expression in terms of the known multiplet that satisfies the desired properties. Then all the other composites appearing in the 4-form $J$ are obtained by successive application of Q-supersymmetry on $C^{ij}$, which ultimately gives us the desired action. The action for pure $\cN=3$ conformal supergravity was constructed in \cite{Hegde:2021rte} using the above mentioned density formula and embedding the $\cN=3$ Weyl multiplet into it, i.e., finding the composites $C^{ij}$ and everything else related to it by supersymmetry in terms of the components of the Weyl multiplet. 
	
	\subsection{Truncation of the Weyl multiplet from $\cN=4$ to $\cN=3$}\label{sec-reduction}
	To make the paper self-contained in this subsection, we review the off-shell reduction of the $\NN=4$ Weyl multiplet to the $\NN=3$ Weyl multiplet as discussed in \cite{Hegde:2021rte}. The independent fields of the $\NN=4$ Weyl multiplet possesses an $SU(4)$ $R$-symmetry appropriate for the $SU(2,2|4)$ algebra. An auxiliary $U(1)$ R-symmetry has been added to the algebra so that the scalar sector can be described by an $SU(1,1)$ valued scalar $\phi_{\alpha}$, where $\alpha=1,2$. It obeys the constraint $\phi_{\alpha}\phi^{\alpha}=1$, where $\phi^\alpha$ is related to $\phi_{\alpha}$ by complex conjugation $\phi^{\alpha}=\eta^{\alpha\beta}\phi_{\beta}^{*}$. The metric $\eta_{\alpha\beta}=\text{diag}(+1,-1)$. The gauge field for the $SU(4)$ $R$-symmetry is $V_\mu{}^I{}_J$ where $I,J=1,\ldots,4$. For performing the supersymmetry reduction of the $\cN=4$ Weyl multiplet, we will decompose the $SU(4)$ index $I$ into $4$ and the $SU(3)$ index $i$ which takes value from $i=1,2,3$. The gauge field $a_{\mu}$ corresponding to the auxiliary $U(1)$-symmetry is composite. It is determined by solving the following constraint in terms of the independent fields of the $\cN=4$ Weyl multiplet.
	\begin{align}\label{phiconstraint}
	\phi^\alpha D_\mu\phi_\alpha=-\frac{1}{4}\bar{\Lambda}^I\gamma_\mu\Lambda_{I},\;\;\;\;\;\;\left(\a =1,2\right).
	\end{align} 
	To truncate this multiplet to the $\cN=3$ Weyl multiplet, we need to set the fourth supersymmetry to zero. Thus, we demand,
	\begin{align}\label{epsilon-reduction}
	\epsilon^4=0=\psi_\mu^4,
	\end{align}
	and follow through the transformation rule of the $\cN=4$ Weyl multiplet to obtain the conditions on the fields. For instance, if we demand consistency of the above condition with the transformation of  gravitino field $\psi_\mu^I$, we get 
	\begin{align}\label{TVreduction}
	T_{ab}{}^{i4}=0=V_\mu{}^4{}_i\;,\; \Lambda_{i}=0\;.
	\end{align}
	In the next step, we identify $\psi_\mu^i$ coming from the $\cN=4$ Weyl multiplet as the gravitino in the $\cN=3$ Weyl multiplet and compare its transformation with \eqref{N3susy} to get,
	\begin{align}
	T_{ab}{}^{ij}&=-\frac{1}{4}\varepsilon^{ijk}T_{abk},\nn\\
	\Lambda_{4}&=\Lambda_L.
	\end{align}
	In arriving at the above expression, we have related the $SU(4)$ Levi-Civita and $SU(3)$ Levi-Civita as $\varepsilon^{ijk4}:=\varepsilon^{ijk}$. In the above equations, we have fields belonging to the $\cN=4$ Weyl multiplet on the L.H.S and on the RHS we have fields that belong to the $\cN=3$ Weyl multiplet. Further, if we use the conditions \eqref{TVreduction} on the transformation of $V_\mu{}^4{}_i$ we get the following condition,
	\begin{align}
	P_a=\varepsilon_{\alpha\beta}\phi^\alpha D_a\phi^\beta=0.
	\end{align}
	We can satisfy this condition by setting the scalars $\phi^\alpha$ to constant values which is consistent with the condition $\phi^\alpha\phi_\alpha=1$. We can either truncate the scalar fields as given below or any other truncation related to it by a rigid $SU(1,1)$ transformation. 
	\begin{align}\label{phireduction}
	\phi_1=1, \phi_2=0.
	\end{align}
	Since the $\cN=3$ Weyl multiplet would belong to an $SU(1,1)$ invariant subsector of the $\cN=4$ Weyl multiplet, the precise details of the scalar field truncation would not matter for the truncation of the $\cN=4$ Weyl multiplet to $\cN=3$. One can also check that the above condition is consistent with the supersymmetry transformations of the scalars $\phi_\alpha$, since $\Lambda_{i}=0$ as shown in (\ref{TVreduction}). From the above truncation of the scalars \eqref{phireduction}, we can also solve the constraint \eqref{phiconstraint} to obtain the gauge field $a_\mu$ of the auxiliary $U(1)$ R-symmetry as,
	\begin{align}\label{amu}
	a_\mu=\frac{i}{4}\bar{\Lambda}_R\gamma_\mu\Lambda_L.
	\end{align}
	Note that since the gauge field $a_{\mu}$ is invariant under the rigid $SU(1,1)$ transformations, the precise details of the reduction of the scalar fields would not affect its expression given in (\ref{amu}). We can now proceed in the similar fashion and infer the full reduction of the $\cN=4$ Weyl multiplet to the $\NN=3$ Weyl multiplet, which can be summarized as follows. On the left hand side we have quantities coming from the $\cN=4$ Weyl multiplet and on the right hand side we have quantities coming from the  $\cN=3$ Weyl multiplet.
	\begin{align}\label{5.12}
	&T_{ab}{}^{ij}= -\frac{1}{4}\varepsilon^{ijk}T_{abk}\;, T_{ab}{}^{4i}=0\;,  \psi_{\mu}^{i}= \psi_{\mu}^{i}\;, \psi_{\mu}^{4}=0\;,\nn\\ &\phi_{\mu}^{i}=\phi_{\mu}^{i}\;, \phi_{\mu}^{4}=0\;, \omega_{\mu}{}^{ab}=\omega_{\mu}{}^{ab}\;, f_{\mu}^{a}= f_{\mu}^{a}\;, b_{\mu}= b_{\mu}\;,a_\mu=\frac{i}{4}\bar{\Lambda}_R\gamma_\mu\Lambda_L\;, \nonumber \\
	&V_{\mu}{}^{4}{}_{4}= \frac{3i}{2}A_{\mu}+\frac{3}{8}\bar{\Lambda}_{R}\gamma_{\mu}\Lambda_{L}\;, V_{\mu}{}^{i}{}_{j}= V_{\mu}{}^{i}{}_{j}-\frac{i}{2}\delta^{i}{}_{j}A_{\mu}-\frac{1}{8}\delta^{i}_{j}\bar{\Lambda}_{R}\gamma_{\mu}\Lambda_{L}\;,\nn\\ 
	&\phi_1= 1\;, \phi_2=0\;, \Lambda_{4}=\Lambda_{L}\;, \Lambda_{i}=0\;, E_{4j}=-\frac{1}{4}E_{j}\;, E_{ij}=0\;,\nonumber \\
	& \chi^{4}_{4j}=\frac{1}{24}\zeta_{j}+\frac{1}{24}E_{k}\Lambda_R\;, \chi^{i}_{jk}=-\frac{1}{16}\varepsilon_{jkm}\chi^{im}-\frac{1}{24}\delta^{i}_{[j}\zeta_{k]}-\frac{1}{24}\delta^{i}_{[j}E_{k]}\Lambda_R\;, \nonumber \\
	&\chi^{4}_{ij}=0\;, \chi^{i}_{4j}=0\;, D^{4i}{}_{4j}=-\frac{1}{48}D^{i}{}_{j}\;, D^{ij}{}_{kl}=\frac{1}{12}\delta^{[i}{}_{[k}D^{j]}{}_{l]}\;, D^{4i}{}_{jk}=0\;, D^{ij}{}_{4k}=0.
	\end{align}
	One can also check that the transformations of the $\cN=3$ Weyl multiplet is reproduced from the transformation rules of the $\cN=4$ Weyl multiplet upon using the above reduction.
\section{The $\cN=3$ vector multiplet and its coupling to conformal supergravity}\label{sec:vector}

In order to study the coupling of $\NN=3$ vector multiplet to conformal supergravity we will perform a supersymmetric truncation of the $\NN=4$ vector multiplet coupled to conformal supergravity, similar in spirit to the truncation of $\NN=4$ Weyl multiplet to the $\NN=3$ Weyl multiplet studied in \cite{Hegde:2021rte} and reviewed in the previous section. The theory of an arbitrary number of $\cN=4$ abelian vector multiplets coupled to conformal supergravity has been studied earlier by de Roo \cite{deRoo:1984zyh}. The vector multiplet consists of a gauge field $\mathcal{A}_{\mu}$, spin-$\frac{1}{2}$ gaugino $\psi_I$ and pseudo-real scalars $\phi_{IJ}$ where $I,J=1,..,4$. In order to perform the truncation of the $\NN=4$ vector multiplet to the $\NN=3$ vector multiplet, we define the following reduction
\begin{equation} \label{eq:N4-N3map}
\begin{aligned}
&\psi_i= \Phi \psi_i\ ;\quad \psi^i=\Phi^{*}\psi^i\ ;&\quad \theta_R=\Phi^* \psi_4\ ;\quad \theta_L=\Phi \psi^4\ ,\\
&\varepsilon_{ijk}\xi^k= \Phi \phi_{ij}\ ;\quad \varepsilon^{ijk}\xi_k= \Phi^* \phi^{ij}\ ;&\quad \xi_i=\Phi^*\phi_{4i}\ ;\quad \xi^i=\Phi \phi^{4i}\ .
\end{aligned}
\end{equation}
The quantities appearing on the LHS are those of the $\NN=3$ theory while the quantities on the RHS are the $\NN=4$ fields. As we can see from above, the $\cN=3$ vector multiplet is not a truncated version of $\cN=4$ vector multiplet but is rather a re-arrangement since none of the $\cN=4$ vector multiplet fields are set to zero in the process. The term $\Phi$ and its complex conjugate appearing above are defined as follows
\begin{align}
\Phi&=\phi^1+\phi^2\;,\; \Phi^*=\phi_1-\phi_2\;,
\end{align}
where $\phi^1$ and $\phi^2$ are the scalar fields appearing in the $\cN=4$ Weyl multiplet defined in section-\ref{sec-reduction}. 

In Table-\ref{Table-vector}, we list the components of the $\cN=3$ vector multiplet along with their $SU(3)$ representations as well as chiral and Weyl weights.
\begin{table}[H]
	\caption{Fields of the $\cN = 3$ vector multiplet}\label{Table-vector}
	\begin{center}
		\begin{tabular}{ | C{2cm}|C{2cm}|C{3cm}|C{2cm}|C{2cm}| }
			\hline
			Fields & Type & $SU(3)$ & $w$ & $c$ \\ \hline
			$\mathcal{A}_{\mu}$ & Boson & \textbf{1} & 0 & 0 \\ \hline
			$\xi_i$ & Boson & \textbf{3} & 1 & -1 \\ \hline 
			$\psi_i$ & Fermion & \textbf{3} & $\frac{3}{2}$ & $\frac{1}{2}$ \\ \hline
			$\theta_L$ & Fermion & \textbf{1} & $\frac{3}{2}$ & $\frac{3}{2}$ \\ \hline
		\end{tabular}
	\end{center}
\end{table}

\noindent 	
The $Q$ and $S$ supersymmetry transformations of the components of the $\NN=3$ vector multiplet coupled to conformal supergravity can be deduced from that of the $\NN=4$ transformation laws \cite{deRoo:1984zyh} along with the truncation defined in \eqref{eq:N4-N3map} as well as the truncation of the Weyl multiplet discussed in section-\ref{sec-reduction}. The $Q$ and $S$ supersymmetry transformation laws for the $\NN=3$ vector multiplet reads as
\begin{align}\label{N3susy}
\delta \mathcal{A}_{\mu}&= \bar{\epsilon}^{i}\gamma_{\mu}\psi_{i}-2\bar{\epsilon}^{i}\psi_{\mu}^j\xi^k \varepsilon_{ijk} - \bar{\epsilon}_{i}\gamma_{\mu}\Lambda_L \xi^i +\thc \nonumber \\
\delta \psi_{i}&= -\frac{1}{2}   \gamma \cdot \mathcal{F}^+ \epsilon_i - 2 \varepsilon_{ijk} \slashed{D} \xi^k \epsilon^j - \frac{1}{4} E_i \xi^k \epsilon_k + \frac{1}{2} \bar{\Lambda}_L \theta_L \epsilon_i\nn \\
&\quad + \frac{1}{2}\gamma_a \epsilon^j \bar{\Lambda}_R \gamma^a \Lambda_L \xi^k \varepsilon_{ijk} + 2 \varepsilon_{ijk} \xi^j \eta^k 
\nonumber \\
\delta \theta_L &= - 2 \slashed{D} \xi^i \epsilon_i -  \gamma^a \bar{\Lambda}_L \gamma_a \Lambda_R \xi^j \epsilon_j + \frac{1}{4} \varepsilon_{ijk} E^i \xi^j \epsilon^k - \bar{\Lambda}_R \psi_i \epsilon^i  - 2 \xi^i \eta_i 
\nonumber \\
\delta \xi_i&=-\bar{\epsilon_i} \theta_R + \varepsilon_{ijk} \bar{\epsilon}^j \psi^k  \nonumber \\
\end{align}
In the transformation law above, the quantity $\mathcal{F}_{ab}^+$ is the self dual component of a modified superconformal field strength defined as
\begin{align}\label{fs}
\mathcal{F}_{ab}^+ = \widehat{F}_{ab}^+ - \frac{1}{4} \bar{\Lambda}_R \gamma_{ab} \theta_R - \frac{1}{2} T_{ab}^i \;\xi_i 
\end{align}	
where $\widehat{F}_{ab}^+$ is the self-dual part of the standard supercovariant field strength corresponding to the gauge field $\mathcal{A}_{\mu}$. Armed with the above dictionary of reduction of the $\NN=4$ vector multiplet to the $\NN=3$ vector multiplet, we look at the equations of the motion of the fields in the $\NN=3$ vector multiplet coupled to conformal supergravity. In order to do this, we start with the equations of motion for the $\NN=4$ vector multiplet coupled to conformal supergravity \cite{deRoo:1984zyh} and use \eqref{eq:N4-N3map} to obtain the corresponding equations of motion for the $\NN=3$ theory. 

The equations of motion of the spin-$\frac{1}{2}$ fermionic field $\psi_I$ in the $\NN=4$ theory will result in two equations of motion, one for $I \equiv i$ and another for $I \equiv 4$ which will give rise to the equations of motion for $\psi_i$ and $\theta_R$ respectively as shown below:
\begin{align}\label{ferm_eom}
&\slashed{D} \psi_i+\frac{1}{2}\bar{\Lambda}_R \psi_i \Lambda_L-\frac{1}{8}E_i \theta_L+\frac{1}{8}\gamma \cdot T_i \theta_L+\frac{1}{8}\chi_{ij}\xi^j +\frac{1}{24}\varepsilon_{ijk}\zeta^j \xi^k=0\ , \nn\\
&\slashed{D}\theta_R -\frac{3}{4}\bar{\Lambda}_R\theta_R \Lambda_L+ \frac{1}{4}\gamma  \cdot \hat{F}^{-} \Lambda_L   -\frac{3}{8}\bar{\Lambda}_L \Lambda_L \theta_L -\frac{1}{8}\gamma \cdot T_i\Lambda_L \xi^i-\frac{1}{8}E_i \psi^i-\frac{1}{8} \gamma \cdot T_i \psi^i\nn \\
&-\frac{1}{12}\zeta^i \xi_i -\frac{1}{8}E^i\xi_i \Lambda_L=0\ .
\end{align}
The equation of motion of $\xi_i$ can be obtained from the equations of motion of $\phi_{IJ}$ (in the $\NN=4$ theory) by fixing $I \equiv 4$ and $J \equiv j$. This yields,
\begin{align}\label{scalareom}
&\square_{c} \xi_j -\frac{1}{4}\bar{\Lambda}_R \slashed{D} \xi_j\Lambda_L  +\frac{1}{4}\widehat{F}^{-}\cdot T_j -\frac{1}{16}\bar{\Lambda}_L \gamma \cdot T_j \theta_L -\frac{1}{8}\xi^i T_i \cdot T_j+\frac{1}{24}\bar{\zeta}_j \theta_R +\frac{1}{16}E_j \bar{\Lambda}_R \theta_R \nn \\
& -\frac{1}{16}\bar{\chi}_{mj}\psi^m -\frac{1}{48}\varepsilon_{jkl}\bar{\zeta}^l \psi^k -\frac{1}{48}D^l_{\ j}\xi_l-\frac{1}{96} \xi_j E^k E_k+\frac{1}{12}\xi_j (\bar{\Lambda}_R \slashed{D}\Lambda_L+\bar{\Lambda}_L \slashed{D} \Lambda_R)\nn \\
&+\frac{1}{12}\xi_j \bar{\Lambda}_R \Lambda_R \bar{\Lambda}_L \Lambda_L=0\ ,
\end{align}
where $\Box_c=D_aD^a$ is the superconformal de-Alembertian.
\section{$\cN=3$ vector multiplet action}\label{sec:embedding}
In this section, we will derive an action for the $\cN=3$ vector multiplet coupled to conformal supergravity. We will embed the vector multiplet in the $\cN=3$ density formula derived in \cite{Hegde:2021rte} and reviewed in section-\ref{density-formula}. However, unlike the Weyl multiplet, $\cN=3$ vector multiplet is on-shell. Since the covariant superform method to obtain the density formula relies on the closure of the superconformal algebra which is now realized up to equations of motion, we would miss the information on terms proportional to equations of motion in the action. Upon adding terms proportional to equations of motion and imposing consistency with the equations of motion given in the previous section, we will obtain the action for $\cN=3$ vector multiplet coupled to conformal supergravity.

As per the density formula reviewed in section-\ref{density-formula}, in order to embed the $\cN=3$ vector multiplet in this density formula, we need to find a suitable composite of the vector multiplet which transforms as a $\mathbf{6}$ representation of $SU(3)$ with Weyl weight $+2$, chiral weight $+2$, is $S$-invariant and satisfies the constraints (\ref{C_constraints}), so that it serves as $C^{ij}$, the lowest component of the abstract multiplet. Such a combination exists up to a complex rescaling as follows,
\begin{align}\label{cdef}
	C^{ij}= \alpha \xi ^i \;\xi^j, 
\end{align} 
where $\alpha$ is an arbitrary complex number. Analogous to the case of pure conformal supergravity action as discussed in \cite{Hegde:2021rte}, when $\alpha$ is real, the density formula leads to a total derivative action. Since we are interested in the supersymmetric completion of the Maxwell's action, we take $\alpha$ to be purely imaginary and set $\alpha=i$. In this section, we are interested in obtaining the supercovariant part of the action which is encoded in the composite $\mathcal{L}$ that appears with the $e^4$ term of the 4-form $J$ in the density formula (\ref{full-density}). And therefore, in this section we will outline the results that are relevant for the computation of $\mathcal{L}$. However, we would like to stress that one can indeed find all the composites needed for the full action including the gravitino terms by following the density formula discussed in section-\ref{density-formula}.

Applying supersymmetry on the above $C^{ij}$ and using (\ref{left_right_Cij}), we get the following composites,
\begin{align}\label{rhodef}
	\hat{\rho}^i&= - 4i \; \xi^i \; \theta_L, \nn  \\
	\rho^i{}_l&= -3i (\xi^i \psi_l-\frac{1}{3}\delta^i_l\; \xi^k \psi_k).
\end{align}
The right-supersymmetry transformation of $\rho^i{}_j$ in \eqref{right_susy_rho} yields the composites $F_{ij}$ and $G_{ab}{}^i$ as given below, 
\begin{align}
	F_{ij}&= 6i \bar{\psi}_i \; \psi_j \nn\\
	G_{ab}^i&= 32i \;\xi^i \mathcal{F}_{ab}^+ + 4i \varepsilon^{ijk} \bar{\psi}_j \gamma_{ab} \psi_k  .
\end{align}
Further, applying supersymmetry on $K_{ij}$ which is obtained from $F_{ij}$ (\ref{KG}) and the composite $\mathcal{G}_{a}{}^{i}{}_{j}\equiv \bar{\Lambda}_{L}\gamma_{a}\rho^{i}{}_{j}-\text{h.c}$, we get:
\begin{align}
	\theta_i&= \frac{i}{2} \gamma \cdot \widehat{F}^+ \psi_i - \frac{i}{4}\gamma \cdot T^j \xi_j \psi_i -\frac{i}{16} \psi_i E_j \xi^j - \frac{i}{16}\psi_j \xi^j E_i +\frac{i}{8}\gamma\cdot T^{j}\Lambda_{R} \xi_{i} \xi_{j}+\frac{i}{8} E^j\Lambda_R \xi_i \xi_j   \nn \\
	&\quad + \frac{i}{2} \bar{\Lambda}_L \theta_L \psi_i -\frac{i}{8} \gamma^{ab} \psi_i \bar{\Lambda}_R \gamma_{ab} \theta_R + i \bar{\Lambda}_R \Lambda_R \xi_i \theta_R   \nn\\
	\tilde{\theta}_i&= 3i \psi_i \xi^j E_j - i \psi_j \xi^j E_i -16i \;\varepsilon_{ijk} \xi^j\slashed{D}\xi^k \Lambda_L - 2i \Lambda_R  E^j  \xi_i \xi_j - 16i \psi_i \bar{\Lambda}_L \theta_L - 8i \bar{\Lambda}_R \Lambda_R \xi_i \theta_R  
\end{align}
Using the expressions for $\theta_{i}$ and $\tilde{\theta}_{i}$ obtained above in (\ref{MN-composites}), we can determine the composite $\mathcal{N}_i$ in terms of the vector multiplet fields as,
\begin{align}
	\mathcal{N}_i &=  \frac{i}{8} \gamma \cdot \widehat{F}^+ \psi_i - \frac{i}{16} \gamma \cdot T^j \xi_j \psi_i - \frac{i}{48} \psi_j \xi^j E_i  -\frac{i}{12} \varepsilon_{ijk} \xi^j\slashed{D}\xi^k \Lambda_L  \nn \\
	&\quad +	\frac{i}{16}\gamma\cdot T^{j}\Lambda_{R} \xi_{i} \xi_{j}   + \frac{i}{48} \Lambda_R \xi_i \xi_j E^j  + \frac{i}{24} \bar{\Lambda}_L \theta_L \psi_i  \nn \\
	&\quad  - \frac{i}{32} \gamma^{ab} \psi_i \bar{\Lambda}_R \gamma_{ab} \theta_R  + \frac{5i}{24} \bar{\Lambda}_R \Lambda_R  \xi_i \theta_R 
\end{align}	
From the right-supersymmetry transformation of the composite $\mathcal{N}_i$ \eqref{3.40}, we can read off $Y$ as follows.
\begin{align}
	Y &= -\frac{3i}{2}\; \widehat{F}^+ \cdot \widehat{F} + \frac{3 i}{2} \widehat{F}^+ \cdot T^i \xi_i - \frac{i}{4} T^i \cdot T^j \xi_i \xi_j  + \frac{i}{96} E^i E^j \xi_i \xi_j - \frac{i}{96} E_i E_j \xi^i \xi^j \nn \\
	&\quad  -\frac{3i}{4}\bar{\psi}_j \slashed{D} \psi^j+ i\bar{\psi}_i \slashed{D} \Lambda_L \xi^i + i \bar{\psi}_i\slashed{D}\xi^i\Lambda_L + \frac{i}{6}\bar{\Lambda}_L\Dslash \psi_i \xi^i + \frac{3i}{4} \bar{\Lambda}_R \gamma \cdot \widehat{F}^+ \theta_R \nn \\
	&\quad - \frac{3i}{32} \bar{\chi}^{ij}\psi_i \xi_j - \frac{i}{32} \varepsilon^{ijk}  \bar{\psi}_i \zeta_j \xi_k +\frac{3i}{32}\bar{\psi}_i \theta_R E^i -\frac{3i}{32}\bar{\psi}_i\gamma\cdot T^i\theta_R \nn \\
	&\quad  {- \frac{i}{48}} \bar{\Lambda}_R \gamma \cdot T^i \theta_R \xi_i {+ \frac{i}{24} \bar{\Lambda}_L \gamma \cdot T_i \theta_L \xi^i } +\frac{5i}{48}\bar{\Lambda}_R \chi^{ij} \xi_i \xi_j +\frac{i}{24}\bar{\Lambda}_L\chi_{ij}\xi^i\xi^j\nn \\
	&\quad + \frac{i}{24} \bar{\Lambda}_L \theta_L \xi^i E_i - \frac{3i}{16}  \bar{\Lambda}_R \theta_R  E^i  \xi_i  -\frac{3i}{8}\bar{\Lambda}_R\psi_i \bar{\Lambda}_L\psi^i  +\frac{i}{16}\bar{\Lambda}_L \Lambda_L \bar{\theta}_L \theta_L\nn\\
	&\quad -\frac{i}{16} \bar{\Lambda}_R \Lambda_R \bar{\theta}_R \theta_R {+\frac{i}{6}}\bar{\Lambda}_L\Lambda_L\bar{\Lambda}_R\psi_i\xi^i + \frac{5i}{12} \bar{\Lambda}_R \Lambda_R \bar{\Lambda}_L \psi^i \xi_i 
\end{align}
And finally using the above expression in (\ref{comp_L}), we can obtain the composite $\mathcal{L}$, which encodes the supercovariant part of the Lagrangian density.	
\begin{align}
	\cL &= - \frac{3i}{2}\; \widehat{F}^+ \cdot \widehat{F} + \frac{3i}{2}\widehat{F} \cdot T^i \xi_i  -\frac{3i}{4}\bar{\psi}_j \slashed{D} \psi^j + i\bar{\psi}_i \slashed{D} \Lambda_L \xi^i + i \bar{\psi}_i\slashed{D}\xi^i\Lambda_L\nn \\
	&\quad  + \frac{i}{6}\bar{\Lambda}_L\Dslash \psi_i \xi^i  + \frac{3i}{4} \bar{\Lambda}_R \gamma \cdot \widehat{F}^+ \theta_R  - \frac{3i}{32} \bar{\chi}^{ij}\psi_i \xi_j - \frac{i}{32} \varepsilon_{ijk}  \bar{\psi}^i \zeta^j \xi^k \nn \\
	&\quad +\frac{3i}{32}\bar{\psi}_i \theta_R E^i -\frac{3i}{32}\bar{\psi}_i\gamma\cdot T^i\theta_R {+ \frac{7i}{48} \bar{\Lambda}_L \gamma \cdot T_i \theta_L \xi^i }  + \frac{7i}{48}\bar{\Lambda}_R \chi^{ij} \xi_i \xi_j \nn \\
	&\quad- \frac{7i}{48} \bar{\Lambda}_R \theta_R  E^i \xi_i  -\frac{3i}{8}\bar{\Lambda}_R\psi_j\bar{\Lambda}_L\psi^j  + {\frac{7i}{12}} \bar{\Lambda}_R \Lambda_R \bar{\Lambda}_L \psi^i \xi_i - \thc
\end{align}
The supercovariant Lagrangian density for a single vector multiplet coupled to conformal supergravity is given as $\mathcal{L}_{V}=-ie\mathcal{L}$. This Lagrangian, however, misses terms proportional to equations of motion, as we discussed earlier. In fact, we can see that the above Lagrangian has no kinetic term for the fields $\theta_L$ and $\xi^i$. To amend this, we need to add terms proportional to equations of motion and, in turn, demand consistency of the Lagrangian with equations of motion.  	

We add kinetic terms of $\theta_L$ and $\xi^i$ by using their equations of motion multiplied by the respective conjugate field with an arbitrary coefficient. These terms are of the form $\bar{\theta}_L(\slashed{D}\theta_R+\ldots)$ and $\xi_i(\square\xi^i+....)$ and their hermitian conjugates. The dots in the brackets indicate the terms from equations of motion obtained in the previous section (\ref{ferm_eom},\ref{scalareom}). Further, we also allow the coefficient of the $\psi_i$ kinetic term to change by adding $\bar{\psi}_j(\slashed{D}\psi^j+\ldots)$ and its hermitian conjugate with an arbitrary coefficient. We would like to fix the coefficients by demanding consistency with equations of motion. But this consistency requirement also forces us to add a term of the form $\xi^i\bar{\Lambda}_L(\slashed{D}\psi_i+\ldots)$ and its hermitian conjugate so that certain unwanted terms do not appear in equations of motion. Finally, all the arbitrary coefficients are determined by using the consistency of the action with the equations of motion, which finally gives us the correct Lagrangian density. With future use in mind, we present this supercovariant part of the Lagrangian density with a generalization to $n$ abelian vector multiplets below.

\begin{align}\label{lag}
e^{-1}\cL_V &=  -\frac{3}{2}\; \widehat{F}^{+I} \cdot \widehat{F}^{J} \eta_{IJ} + 6\; \xi^{I}_i \square \xi^{Ji} \eta_{IJ}  + {3}\;\widehat{F}^I \cdot T^i \xi_i^J \eta_{IJ} - \frac{1}{8}D^i_{\ j}\xi_i^I \xi^{Jj} \eta_{IJ} \nn \\
	&\quad -\frac{3}{4} T_i \cdot T_j \xi^{Ii} \xi^{Jj} \eta_{IJ} -\frac{1}{16} E^i E_i\xi^{Ij} \xi_j^{J} \eta_{IJ} - {3}\bar{\psi}_j^{I} \slashed{D} \psi^{Jj}\eta_{IJ} - 3 \bar{\theta}_L^{I} \slashed{D} \theta_R^{J} \eta_{IJ}  \nn \\
	&\quad  + \bar{\psi}_i^I \slashed{D} \Lambda_L \xi^{Ji} \eta_{IJ} + \bar{\psi}_i^I\slashed{D}\xi^{J i}\Lambda_L \eta_{IJ} - \bar{\Lambda}_L\Dslash \psi_i^I \xi^{J i} \eta_{IJ} + \frac{3}{2} \bar{\Lambda}_R \gamma \cdot \widehat{F}^{+I} \theta_R^J \eta_{IJ} \nn \\
	&\quad - \frac{3}{4} \bar{\chi}^{ij}\psi_i^I \xi_j^J \eta_{IJ} - \frac{1}{4} \varepsilon_{ijk}  \bar{\psi}^{Ii} \zeta^j \xi^{J k} \eta_{IJ}  +\frac{1}{2} \bar{\theta}_L^I \zeta^i \xi_i^J \eta_{IJ} +\frac{3}{4}\bar{\psi}_i^I \theta_R^J E^i \eta_{IJ}   \nn \\
	&\quad + \frac{3}{4}\bar{\psi}_i^I\gamma\cdot T^i\theta_R^J \eta_{IJ} + \frac{3}{4} \bar{\Lambda}_L \gamma \cdot T_i \theta_L^I \xi^{Ji} \eta_{IJ} +\frac{3}{4} \bar{\Lambda}_L \theta_L^I  E^i \xi_i ^J \eta_{IJ} -\frac{3}{2}\bar{\Lambda}_R \slashed{D} \xi_j^I \Lambda_L \xi^{Jj} \eta_{IJ}\nn \\
	&\quad  +\frac{1}{2}\xi_j ^I\xi^{Jj}(\bar{\Lambda}_R \slashed{D}\Lambda_L  +\bar{\Lambda}_L \slashed{D} \Lambda_R) \eta_{IJ} - {3}\; \bar{\Lambda}_R\psi_j^I\bar{\Lambda}_L\psi^{J j} \eta_{IJ} +\frac{9}{4}  \bar{\Lambda}_L \theta_L^I\bar{\Lambda}_R \theta_R^J \eta_{IJ}  \nn \\
	&\quad + \frac{9}{8} \bar{\Lambda}_L \Lambda_L \bar{\theta}_L^I \theta_L^J \eta_{IJ}  + \frac{1}{2} \xi_j^I  \xi^{J j} \bar{\Lambda}_R \Lambda_R \bar{\Lambda}_L \Lambda_L \eta_{IJ} + \text{h.c}
\end{align}

The indices $I, J = 1,.....,n$ label the vector multiplets. As we will discuss in the next section, we will need three compensating vector multiplets to obtain pure Poincar\'e supergravity. The kinetic term of the compensating multiplets should come with the wrong sign so that upon using the Poincar{\'e} gauge fixing condition, the kinetic term of the physical fields such as graviton, graviphoton, gravitino, etc comes with the right sign. However, if we are interested in studying coupling of $n_v$ vector multiplets to Poincar{\'e} supergravity, we would need to consider $3+n_v$ vector multiplets coupled to conformal supergravity, where the sign of the kinetic terms for 3 of the vector multiplets should come with a wrong sign and for the remaining $n_v$ vector multiplets, it should come with the correct sign.  Therefore one would need to take  $\eta_{IJ}=\text{diag}\{-1,-1,-1,1,\ldots,1\}$ in the above Lagrangian. In this paper, however, we limit ourselves to $3$ vector multiplets coupled to conformal supergravity and take $\eta_{IJ}=\text{diag}(-1,-1,-1)$, because we are interested in studying pure $\mathcal{N} = 3$ Poincar{\'e} supergravity. Note that in the above Lagrangian, the fields $D^{i}{}_{j}$, $\chi^{ij}$ and $\zeta^{i}$ belonging to the $\cN=3$ Weyl multiplet appear as Lagrange multiplier and imposes 8 bosonic and 36 fermionic constraints on the vector multiplet fields. This observation will play an important role in our discussion in the next section.

	\section{Gauge fixing and pure Poincar{\'e} supergravity}\label{sec:gaugefixing}
Now that we have obtained the superconformal action, we are in a position to write the action for the super-Poincar\'e theory by breaking the superfluous symmetries in the superconformal theory. In this section, we will discuss the gauge fixing of the superconformal group to the super-Poincar\'e group, and subsequently, we will write down the action in the super-Poincar\'e theory. 

Starting with special conformal symmetry, which can be broken by simply turning off the only gauge field on which they act, the dilatation gauge field $b_\m$.
\begin{align}\label{Kgauge}
K - \text{gauge}: \;\;\;\;\;\;\;b_\mu = 0
\end{align}
Furthermore, extra 18 (bosonic) + 48 (fermionic) degrees of freedom are required to gauge fix dilatation (1B), $SU(3)_R$ (8B), $U(1)_R$  (1B), and $S$-supersymmetry (12F), as well as to satisfy the constraints imposed by the Lagrange multipliers $D^i{}_j$ (8B), $\chi_{ij}$ (24F), and $\zeta_i$ (12F)\footnote{The numbers inside the brackets denote the degrees of freedom required for the purpose and ``B'' or ``F'' denotes bosonic or fermionic.}. We know that one $\NN=3$ vector multiplet has the fields $\xi_{i}$ (6B), $\psi_{i}$ (12F) and $\theta_L$ (4F) which are relevant for providing 6 (bosonic)+ 16 (fermionic) degrees of freedom for the purpose. Thus, we need at least three vector multiplets as compensators to get the super-Poincar\'e theory. The  vector gauge fields coming from these compensating multiplets would be a part of the super-Poincar{\'e} theory and would be known as graviphotons.

The dilatation symmetry can be broken by imposing the following condition on the scalar
field of the vector multiplet $\xi_i$ :
\begin{align}\label{Dgauge}
D - \text{gauge}: \;\;\;\;\;\;\;\xi^{I i} \xi^{J}_i \eta_{IJ} = - \frac{1}{\kappa^2}\ , 
\end{align}
where $\kappa$ is a dimensionfull constant of mass dimension $-1$ that enters the theory as a result of the breaking of the dilatation symmetry. In order to break $S$-supersymmetry, we choose the following gauge condition,
\begin{align}
S- \text{gauge}:\;\;\;\;\;\;\; \xi_i^I \theta_L^J \eta_{IJ} = 0\ .
\end{align}
\noindent
At this point, the Lagrangian (\ref{lag}) is still invariant under local $SU(3)_R\times U(1)_R$ transformation. However, instead of imposing suitable gauge fixing conditions on the scalars for breaking these symmetries, we note that we can write the bosonic part of the action in terms of an 
$SU(3)\times U(1)$ invariant object $M^{IJ} \equiv  \kappa^2 \xi^{I}_i  \xi^{Ji}$ which satisfy the following conditions
\begin{align}\label{MProp}
M^\dagger = M  \;\;\quad ;\;\;\; Tr (M \eta) &= -1 \;\;\quad ;\;\;\; M^\dagger \eta M = - \frac{1}{3} M
\end{align}
For the minimal case with only 3 coupled vector multiplets, the above conditions imply that 
\begin{align}\label{M_minimal}
M^{IJ} = - \frac{1}{3} \eta^{IJ}\ . 
\end{align}
\noindent
We now focus on the field equations of the Weyl multiplet fields that appear as Lagrange multiplers in the vector multiplet action $(D^i\;_j, \chi_{ij}, \zeta_i)$. 
\begin{itemize}
	\item Field equations of $D^i\;_j$:
	\begin{align}\label{Dijeom}
	(\xi^{I i} \xi^{J}_j \eta_{IJ} )_\textbf{8} =0
	\end{align}
	\item Field equations of $\chi_{ij}$:
	\begin{align}\label{chieom}
	\psi^I_{( i} \xi^{J}_{j)} \eta_{IJ}  =0
	\end{align}
	\item Field equations of $\zeta_{i}$:
	\begin{align}\label{zetaeom}
	\xi_i^I \theta_L^J \eta_{IJ} - \frac{1}{2} \varepsilon_{ijk} \xi^{Ij} \psi^{Jk} \eta_{IJ} = 0
	\end{align}
	
\end{itemize}
After taking S-gauge condition into consideration, the field equations \eqref{chieom} and \eqref{zetaeom} together implies:
\begin{align}\label{Sgauge}
\psi^I_{ i} \xi^{J}_{j} \eta_{IJ}  =0
\end{align}
Note that,
\begin{align}\label{threecross}
{\xi^{I i} \xi^{J}_j \eta_{IJ}}  &= {(\xi^{I i} \xi^{J}_j \eta_{IJ})_{\textbf{8}}} + {\frac{1}{3} \delta_j^i ( \xi^{I k} \xi^{J}_k \eta_{IJ} )} = - \frac{1}{3\kappa^2} \delta_j^i 
\end{align}
where we have used $D$-gauge (\ref{Dgauge}) and the $D^i\;_j$ field equation (\ref{Dijeom}) to reach the above form \eqref{threecross}.

The auxiliary fields $T_{ab}^i$ and $E_i$ appear quadratically in action and, therefore, can be expressed in terms of their field equations as
\begin{align}
E_i &= - 6 \bar{\psi}_i^I \theta_R^J \eta_{IJ} \;  , \\
T_{ab}^j \xi_j^I \xi^J_{ i} \eta_{IJ}  &= 2 \widehat{F}_{ab}^{+I} \xi^J_{ i} \eta_{IJ}  - \frac{1}{2} \bar{\Lambda}_R \gamma_{ab} \theta_R^I \xi^J_{ i} \eta_{IJ}\ . 
\end{align}
Similarly, the equation of motion for the $SU(3)$ and $U(1)$ fields are given as
\begin{align}\label{sueom}
\mathcal{V}_\mu^i\;_j &= \frac{3}{2} \xi^{I i} \overleftrightarrow{\partial}_\mu \xi^J_j \eta_{IJ} - \text{trace} +\text{fermions}, \nn \\
A_\mu &= \frac{i}{2} \xi_i^I \overleftrightarrow{\partial}_\mu \xi^{Ji} \eta_{IJ} + \text{fermions}.
\end{align}

It must be noted that, for the case of pure Poincar{\'e} supergravity, when we have only three vector multiplets as compensators, then the dilatation gauge fixing condition, the constraints imposed by the Lagrange multipler $D^{i}{}_{j}$ together with a suitable $SU(3)_R\times U(1)_R$ gauge fixing condition would imply that all the scalar fields $\xi_{i}^{I}$ take constant values. The constant value would be determined by the exact nature of $SU(3)_R\times U(1)_R$ gauge fixing condition. However, irrespective of the constant values taken by the scalar fields, the $SU(3)_{R}\times U(1)_{R}$ invariant object $M^{IJ}$ constructed out of them is given by (\ref{M_minimal}). This combined with (\ref{Sgauge}) and (\ref{zetaeom}) implies $\theta_L = 0 = \psi_i$. 
Thus, for the minimal case, we get,
\begin{align}\label{LOaux}
	\cV_\m^i{}_j &= 0 \;\;\;,\;\; A_{\m} = 0 \ ,\nn \\
		E_i &= 0 \;\;\;,\;\;	T_{ab}^i  = - 6 \kappa^2\widehat{F}_{ab}^{+I} \xi^{J i} \eta_{IJ}
\end{align}

However, it is important to note that in matter coupled supergravity theories where we have $n_{v}$ additional vector multiplets,  $\xi_i^I$ will no longer be constant. For $(3+ n_v)$ vector multiplets coupled to conformal supergravity, we will have $6(3+n_v)$ scalars, out of which only 18 will be fixed by the gauge fixing conditions leaving behind $6n_v$ physical real scalar ($3n_v$ complex scalar) fields in the matter coupled Poincar{\'e} supergravity theory, which will be encoded in the matrix valued scalar fields $M^{IJ}$ obeying (\ref{MProp}).

\noindent
After imposing all the field equations and gauge fixing conditions on the Lagrangian for the compensating vector multiplets coupled to conformal supergravity,  we obtain the the Lagrangian for pure Poincar\'e supergravity as
\begin{align}\label{vPLag}
e^{-1}\cL_0 &= - \frac{1}{\kappa^2} R + \frac{3}{2}\; \widehat{F}^{+I} \cdot \widehat{F}^{J} \eta_{IJ} +  \text{fermions} + \text{h.c}\ .
\end{align}
The supersymmetry preserved by the above action has to be redefined from the Q-supersymmetry ($\delta_Q$) defined in the superconformal theory. This is because the $K$-gauge and the $S$-gauge condition break $Q$-supersymmetry and hence the unbroken supersymmetry for the super-Poincar{\'e} theory has to be redefined by adding field dependent $K$ and $S$ transformations which preserve the gauge fixing condition. And hence, the Poincar{\'e} supersymmetry transformations $\delta_Q^P$ that is preserved by the super-Poincar{\'e} theory is given as
\be
\delta_Q^{P}(\epsilon_{i}) = \delta_Q(\epsilon_{i}) + \delta_K(\Lambda_{K\mu})+ \delta_S({\eta_i})\ ,
\ee
where,
\begin{align}
\Lambda_{K\mu} &= - \frac{1}{2} \bar{\epsilon}^i \phi_{\mu i} + \text{h.c.}\ ,\nn \\ 
\eta_i &= \frac{3}{4} \bar{\theta}_L^I \gamma^a \theta_R^J \eta_{I J} \gamma_{a} \epsilon_{i} +  \frac{3}{2} \varepsilon_{ijk} \bar{\psi^{kI}} \gamma^a \theta_L^J \eta_{I J} \gamma_{a} \epsilon^{j} -  \frac{1}{2} \bar{\Lambda}_L^I \gamma^a \Lambda_R^J   \gamma_{a}\epsilon_{i}
\end{align}
The bosonic part of the Lagrangian \eqref{vPLag} is given by an Einstein-Maxwell theory for a set of three graviphotons ($3\leq I\leq 1$) coupled to gravity. The bosonic equations of motion are given by the following Einstein-Maxwell's equations:
\begin{align}\label{Ein_Max_eom}
	R_{\m\n} - \frac{1}{2} g_{\m\n} R &= \frac{3}{4} \kappa^2 \left( 2 g^{\r \s} \widehat{F}_{\m \r} \widehat{F}_{\n \s} - \frac{1}{2} g_{\m \n} \widehat{F} \cdot \widehat{F} \right)\ . \nonumber \\
\mathcal{D}_{a} \widehat{F}^{abI} &= 0
\end{align}
The Bianchi identity is given as:
\begin{align}
\mathcal{D}_{a}\widetilde{\widehat{F}}^{abI}=0
\end{align}
where our convention for dual tensor is given in appendix-\ref{appendix-conventions}. The Maxwell's equation together with the Bianchi identity implies:
\begin{align}\label{Max}
\mathcal{D}_{a}\widehat{F}^{+abI}=0=\mathcal{D}_{a}\widehat{F}^{-abI}\;,
\end{align}
In the above equations of motion, we have written the Maxwell's equation and the Bianchi identities in terms of the covariant derivative $\mathcal{D}_a$ which is covariant only w.r.t the local Lorentz transformations\footnote{This descends from the superconformal theory where the derivative $\mathcal{D}_{a}$ is defined to be covariant w.r.t all the standard gauge transformations (dilatation, local Lorentz, $SU(3)\times U(1)$ R-symmetry).}. However, we can convert it into the covariant derivative $\nabla_{\mu}$ which is also covariant w.r.t general coordinate transformation as we explain below.

Recall, that in order to construct a theory of conformal supergravity from superconformal gauge theory, we need to impose the following constraint on the curvature corresponding to the local translation $(P_a)$, so that local translation is realized as general coordinate transformation:
\begin{align}
R(P)_{\mu\nu}^{a}=2\mathcal{D}_{[\mu}e_{\nu]}^{a}+\text{fermions}=0
\end{align}
Let us further define:
\begin{align}\label{christoff}
\mathcal{D}_{(\mu}e_{\nu)}^{a}\equiv\gamma_{\mu\nu}^{a}\;,\; \Gamma_{\mu\nu}^{\rho}\equiv e^{\rho}_{a}\gamma_{\mu\nu}^{a}
\end{align}
And therefore (up to fermions)
\begin{align}\label{covgct}
\nabla_{\mu}e_{\nu}^{a}\equiv \mathcal{D}_{\mu}e_{\nu}^{a}-\Gamma_{\mu\nu}^{\rho}e_{\rho}^{a}=\mathcal{D}_{[\mu}e_{\nu]}^{a}=0
\end{align}
The derivative $\nabla_{\mu}$ defined above is covariant w.r.t all the standard gauge transformations as well as general coordinate transformation with the quantity $\Gamma_{\mu\nu}^{\rho}$ defined above playing the role of the Christoffel connection. Under this covariant derivative, the vielbein $e_{\mu}^{a}$ and consequently the metric $g_{\mu\nu}$ is covariantly constant as it should be. Thus, one can re-write the Maxwell's equation appearing in (\ref{Max}) as\footnote{By following the sequence of steps: $\mathcal{D}_{a}F^{ab}=e^{\mu}_{a}\mathcal{D}_{\mu}F^{ab}=e^{\mu}_{a}\nabla_{\mu}F^{ab}=\nabla_{\mu}(e^{\mu}_{a}F^{ab})=(\nabla_{\mu}F^{\mu\nu})e_{\nu}^{b}=0$.}
\begin{align}\label{Max1}
\nabla_{\mu}\widehat{F}^{+\mu\nu I}=0=\nabla_{\mu}\widehat{F}^{-\mu\nu I}
\end{align}

\section{Higher derivative deformation}\label{HDD}

In the previous section, we obtained the pure $\cN=3$ Poincar{\'e} supergravity action which is second order in derivatives. The goal of this section is to construct a higher derivative deformation to the leading order piece obtained in \eqref{vPLag}. In order to obtain such a deformation in the superconformal set-up one needs to do the following:
\begin{enumerate}
\item
Add the pure $\cN=3$ conformal supergravity action, constructed purely out of the Weyl multiplet, to the action for the vector multiplets coupled to conformal supergravity.
\item
Impose the Poincar{\'e} gauge fixing condition discussed in the previous section.
\item
Eliminate the auxiliary fields $T_{abi}$, $E_{i}$, $D^{i}{}_{j}$, $V_{\mu}{}^{i}{}_{j}$ and $A_{\mu}$. We will see later that the elimination of these fields systematically lead to a derivative expansion of the action. We will also see that the truncation of the action at the fourth order in derivatives is a consistent truncation.
\end{enumerate}
The pure $\cN=3$ conformal supergravity action has been obtained in \cite{Hegde:2021rte}\footnote{We have corrected minor typos in the original paper \cite{Hegde:2021rte}. The negative sign of the term 
	$T^{abi} D_a D^c T_{bci}$ apperaring in the Pontryagin and Weyl square Lagrangian has been corrected to positive. We trace the error back to an appendix and have correct this in the new arXiv version of \cite{Hegde:2021rte}.}. As discussed above, as a first step, we add this action to the action for the vector multiplets coupled to conformal supergravity. For the sake of brevity, we will only consider the bosonic part. Since we are interested in pure supergravity, as we did in the previous section, we will consider only the coupling of 3 compensating vector multiplets to conformal supergravity labelled by the index $I$ with $\eta_{IJ}=\text{diag}(-1,-1,-1)$.
\begin{align}\label{LVCSG}
	\mathcal{L} = \mathcal{L}_V + \lambda \; \mathcal{L}_{CSG}\ ,
\end{align}
where $\mathcal{L}_{V}$ is given in (\ref{lag}) and
\begin{align}
	\label{eq:LCSG}
	\mathcal{L}_{CSG} &= 24R(M)^{abcd}R(M)^+_{abcd}  +48R(V)^+{}^j{}_i\cdot R(V)^+{}^i{}_j -144 R(A) \cdot R^+(A)\nn \\
	&\quad -6R(V){}^j{}_i\cdot T^iE_j +12iR(A)\cdot T^iE_i +\frac{3}{16}T^i\cdot T^jE_iE_j +\frac{1}{48}D^i\;_j D^j\;_i
	\nn\\
	&\quad +\frac{3}{2} E^i D^a D_a E_i + 24 T^{abi} D_a D^c T_{bci} - \frac{1}{128}E_iE^iE_jE^j  + \frac{3}{8} \Big(T^i \cdot T^j \Big) \Big(T_i \cdot T_j \Big)
	\nonumber \\ &\quad + \text{h.c.}
\end{align}
Here $R(M)$, $R(V)$ and $R(A)$ are the fully supercovariant curvatures for the local Lorentz transformation (M), $SU(3)$ R-symmetry (V) and $U(1)$ R-symmetry (A) respectively. Their expressions are given in Appendix-\ref{trans_dep} along with the curvature constraints and the Bianchi identities that they satisfy. The curvature $R(M)_{\mu\nu}{}^{ab}$ becomes the Weyl tensor $C_{\mu\nu}{}^{ab}$ (defined in \ref{Weyl}) upon using the K-gauge condition (\ref{Kgauge}) and inserting the composite expression for the dependent K-gauge field $f_{\mu}^{a}$, which upon using the K-gauge condition is given in terms of the Ricci tensor and Ricci scalar as shown below (See Appendix-\ref{appendix-conventions} for our convention of Riemann tensor, Ricci tensor and Ricci scalar):
\begin{align}
f_{\mu}^{a}=\frac{1}{2}R_{\mu}{}^{a}-\frac{1}{12}Re_{\mu}^{a}
\end{align}
The covariant derivative $D_{a}$ appearing in the action (\ref{eq:LCSG}) is the fully supercovariant derivative. However, if one is only interested in the bosonic part, one may replace it by $D_{a}=\mathcal{D}_{a}+\text{K-covariantization}$. where $\mathcal{D}_{a}$, as discussed in the previous section, is covariant w.r.t all the standard gauge transformations (dilatation, local Lorentz transformation and R-symmetry). The parameter $\lambda$ appearing in (\ref{LVCSG}) is a dimensionless parameter which, as we will see, will control the derivative expansion of the action. Hence, we will also refer to it as the control parameter.

As a second step, we need to impose the Poincar{\'e} gauge fixing conditions, which are the same as discussed in the previous section. As a third step we would need to eliminate the auxiliary fields from their equations of motion. In the absence of the pure conformal supergravity action ($\mathcal{L}_{CSG}$), the field $D^{i}{}_{j}$ appeared as a Lagrange multiplier that imposed the constraints (\ref{Dijeom}) on the vector multiplet scalars. However, in the presence of the conformal supergravity action, the variation of the action (\ref{LVCSG}) w.r.t $D^{i}{}_{j}$ leads to the following equation:
   \begin{align}\label{HDijeom}
		\left[ \xi_i^I \xi^{Jj} \eta_{IJ}\right]_{\textbf{8}} &= \frac{\lambda}{3} D^j\;_i\ . 
	\end{align}
	Taking into account the dilatation gauge fixing condition (\ref{Dgauge}), we can re-write the above equation as
	\begin{align}\label{D_eom}
		\xi_i^I \xi^{Jj} \eta_{IJ} &= - \frac{1}{3 \kappa^2}\delta_i^j + \frac{\lambda}{3} D^j\;_i\ . 
	\end{align}
	Note the appearance of the field $D^{i}{}_{j}$ in the above equation. In the absence of the conformal supergravity action, we really do not worry about the expression for the field $D^{i}{}_{j}$ in terms of the physical fields since the expression in the action involving $D^{i}{}_{j}$ vanishes due to the constraint (\ref{Dijeom}). However, this is not true once we add the conformal supergravity action. Thus, we need to know the expression which would determine $D^{i}{}_{j}$. In this context, the equation of motion for the vector multiplet scalars (\ref{scalareom}) is useful. It is instructive to put the label $I$ on the equations. The bosonic part is given as:
	\begin{align}\label{xi_eom}
		\square_{c} \xi^{Ii} + \frac{1}{4} \widehat{F}^I \cdot T^i - \frac{1}{48} D^i{}_j \xi^{Ij} - \frac{1}{8} T^i \cdot T^j \xi_j^I - \frac{1}{96} E^j E_j \xi^{Ii} = 0 
	\end{align}
 In order to extract $D^{i}{}_{j}$ from the above equation, one needs to contract $\xi^{J}_{k}\eta_{IJ}$ to the above equation and take the projection on the $\textbf{8}$ irrep of $SU(3)$. We get:
	\begin{align}\label{Dij}
	D^{i}{}_{j}&=-144\kappa^2\left(\xi^{J}_{j}\square_{c}\xi^{Ii}\eta_{IJ}\right)_\textbf{8}-36\kappa^2\left(\hat{F}^{I}\cdot T^{i}\xi^{J}_{j}\eta_{IJ}\right)_{\textbf{8}}+18\kappa^2\left(T^{i}\cdot T^{k}\xi^{I}_{j}\xi^{J}_{k}\eta_{IJ}\right)_{\textbf{8}}\nonumber \\
	& +\frac{\lambda\kappa^2}{2}E^{k}E_{k}D^{i}{}_{j}+\lambda\kappa^2\left(D^{i}{}_{k}D^{k}{}_{j}\right)_{\textbf{8}}
	\end{align}
The $T_{abi}$ and $E_{i}$ equations of motion are also modified as follows:
\begin{itemize}
\item Field equations for $E_{i}$:
	\begin{align}\label{E_eom}
		E_i &= - \lambda \kappa^2 \Big( 24 \square_{c} E_i - \frac{1}{4} E_i E_j E^j + 48 R(V)^j\;_i \cdot T_j - 96 R(A) \cdot T_i + 3 T_i \cdot T_j E^j \Big)  \nn\\
		&\equiv -\lambda \kappa^2 \mathcal{P}_i \ .
	\end{align}
	\item Field equations of $T^i_{ab}$:
	\begin{align}\label{T_eom}
		T_{ab}^j \xi_j^I \xi_i^J \eta_{IJ} = 2 \widehat{F}_{ab}^{+I} \xi_i^J \eta_{IJ} + \lambda \mathcal{M}_{ab i}^+ \ ,
	\end{align}
	where,
	\begin{align}\label{MabDef}
		\mathcal{M}_{ab i}  =\;&  32 D_aD^c T_{bc i} - 4 R(V)_{ab}{}^j{}_i E_j + 8i R(A)_{ab} E_i \nn \\
		& + \frac{1}{4} T_{ab}^j E_j E_i + T_{ab}^j T_i \cdot T_j\ . 	
	\end{align}
	\end{itemize}
	Using the equations of motion of $E_{i}$ and $T_{abi}$, one may simplify the equations of motion determining $D^{i}{}_{j}$ (\ref{Dij}) as:
	\begin{align}\label{hdij}
	D^i{}_j =& - 144 \kappa^2 \left[ \xi_j^I \Box_{c} \xi^{J i} \eta_{I J} \right]_{\textbf{8}}
	+ \lambda \kappa^2 \left[ D^i{}_k  D^k{}_j \right]_{\textbf{8}} + 18 \lambda \kappa^2 \left[ T^i \cdot \mathcal{M}_j \right]_{\textbf{8}} \nn \\
	& + \frac{1}{2} \lambda^3 \kappa^6 D^i{}_j \mathcal{P}^2\ .
\end{align}
The equations of motion for the R-symmetry gauge fields $A_{\mu}$ and $V_{\mu}{}^{i}{}_{j}$ are also modified as follows:
\begin{itemize}
	\item Field equations of $A_{\mu}$: 
	\begin{align}\label{hAu}
		A_\mu&=-\frac{i}{2}\kappa^2\xi^i_I\overset{\leftrightarrow}{\partial_\mu}\xi_i^I-\frac{i\lambda\kappa^2}{3}V_\mu^j{}_iD^j_i+\lambda \kappa^2 \Sigma_\mu\ ,
	\end{align}
	where $\Sigma_a\equiv e^{\mu}_{a}\Sigma_{\mu}$ is given by
	\begin{align}\label{sigmaAdef}
		\Sigma_a&=12D^b R(A)_{ab}-D^b\left(T_{ab}^iE_i\right)-\frac{i}{8}E^iD_a E_i-2i(D_c T^{bci})T_{abi}+\thc\ .
	\end{align}
	\item Field equations for $V_\mu^i{}_j$:
	\begin{align}\label{hVij}
		V_\mu^i{}_j=-\frac{3\kappa^2}{2}\xi_i^I\overset{\leftrightarrow}{\partial_\mu}\xi^j_I+2i\lambda\kappa^2A_\mu D^i_j+\lambda\kappa^2\Sigma_\mu^i{}_j,
	\end{align}
	where,
	\begin{align}\label{sigmaVdef}
		\Sigma_a{}^i{}_j&=24D^b R(V)_{ab}{}^i{}_j-3D^b\left(T_{ab}^iE_j\right)+\frac{3}{4}D_a E_j E^i+12T_{abj}D_c T^{bci}-\text{h.c} - \text{trace}.
	\end{align}
\end{itemize}
\noindent
It is evident from the equations of motion that one can obtain the solutions to the auxiliary fields (schematically denoted as $\Phi$ ) in powers of the control parameter $\lambda$ :
\be
\Phi = \Phi^{(0)} + \lambda \Phi^{(1)} + ......
\ee

\noindent
One can also see from the equations of motion  (\ref{E_eom}, \ref{T_eom}, \ref{hdij}, \ref{hAu}, \ref{hVij}) of the auxiliary fields that a factor of $\kappa^2$ accompanies every power of $\lambda$ and hence an expansion in $\lambda$ controls the derivative expansion of the action. From the following argument we show that the zeroth order solutions for the auxiliary fields are sufficient if we are interested in expanding the action up to fourth order in derivatives.

\bea
S [\Phi^{(0)} + \lambda \Phi^{(1)}] &= S_V [\Phi^{(0)} + \lambda \Phi^{(1)}] + \lambda S_{CSG}[\Phi^{(0)} + \lambda \Phi^{(1)}], \nn \\
&=S_V [\Phi^{(0)}] + \lambda \left( \frac{\delta S_V}{\delta \Phi} \right)_{\Phi^{(0)}} \cdot \Phi^{(1)} + \lambda S_{CSG} [\Phi^{(0)}]
\eea

\noindent
Expanding the action to first order in $\lambda$ is adequate because we want to expand the action up to fourth order in derivatives. Furthermore, $\Phi^{(0)}$ is the leading order solution of the auxiliary fields and hence it satisfies $\left(\frac{\delta S_V}{\delta \Phi}\right)_{\Phi_0}=0$. Thus the expansion of the action up to $\mathcal{O}(\lambda)$ or equivalently up to fourth order in derivatives is given by

\be
S [\Phi^{(0)} + \lambda \Phi^{(1)}]= S_V [\Phi^{(0)}] + \lambda S_{CSG} [\Phi^{(0)}]	
\ee
\noindent 
Hence it is sufficient to find the leading order solutions for the auxiliary fields in order to expand the action up to fourth order in derivatives. The leading order solution for all the auxiliary fields except $D^{i}{}_{j}$ is given in (\ref{LOaux}). One can also check from the equation determining $D^{i}{}_{j}$ (\ref{hdij}) that the leading order solution is given as:
\begin{align}\label{L_eom}
	D^i{}_j &=- 144 \kappa^2 \left[ \xi_j^I \Box_{c} \xi^{J i} \eta_{I J} \right]_{\textbf{8}}=0
\end{align}
The above expression vanishes because of the following reason. In the case of pure supergravity, as explained in the previous section, at the leading order the vector multiplet scalars take constant values. Hence $\xi_j^I \Box_{c} \xi^{J i} \eta_{I J}\sim f^{a}_{a}\xi_j^I \xi^{J i} \eta_{I J}\sim R \delta^{i}_{j}$. Therefore it vanishes when we project it on the $\textbf{8}$ irrep of SU(3). Plugging the leading order solutions for the auxiliary fields in \eqref{eq:LCSG}, we obtain the expansion of the $\cN=3$ Poincar{\'e} supergravity action up to fourth order in derivatives as
\begin{align}
	\label{eq:higherderivativeaction}
	e^{-1} \cL = \cL_0 + \lambda\kappa^2 \cL_1\ ,
\end{align}
where the leading piece is
\begin{align}
	\cL_0 = -\frac{2}{\kappa^2} R + \frac{3}{2} \widehat{F} \cdot \widehat{F}\ ,
\end{align}
and the four derivative deformation is given as,
\begin{align}
	\cL_1 = \; &\frac{24}{\kappa^2} C^{\m \n \r\s}C_{\m\n\r\s} + 288  R_{\m\n} \widehat{F}^{+I \m}{}_\r \widehat{F}^{-J \r \n} \eta_{IJ} -576 \nabla_\m\widehat{F}^{+I\m \n} \nabla^\r \widehat{F}^{-J}_{\r \n} \eta_{IJ} \nn \\
	& + 108 \kappa^2(\widehat{F}^{+I}\cdot \widehat{F}^{+K})(\widehat{F}^{-J}\cdot \widehat{F}^{-L})\eta_{KL} \eta_{IJ}\ .
\end{align}
The indices $I, J = 1,2,3$ label the compensating vector multiplets with $\eta_{IJ}=diag(-1,-1,-1)$. In order to expand the Lagrangian to next order in derivatives, we would need to find the corrections to the auxiliary field solutions. The first order corrections to the auxiliary fields are obtained as follows. First, we need to obtain the first order correction ($\mathcal{O}(\lambda)$) to the vector multiplet scalars by substituting the leading order expression for $D^{i}{}_{j}$ in equation (\ref{D_eom}). However since $D^{i}{}_{j}$ vanishes at leading order, we find that the vector multiplet scalars do not receive any correction at the first order. And therefore, the first order corrections to the auxiliary fields are obtained by simply putting the leading order solutions into the $\mathcal{O}(\lambda)$ term appearing in the auxiliary field equations (\ref{E_eom}, \ref{T_eom}, \ref{hdij}, \ref{hAu}, \ref{hVij})). For instance, the $\mathcal{O}(\lambda)$ correction to the $SU(3)$ gauge field is given as:
\begin{align}
V_{\mu}^{(1)}{}^{i}{}_{j}=\lambda\kappa^2 \Sigma_{\mu}^{(0)}{}^{i}{}_{j}
\end{align}
where $\Sigma_{\mu}^{(0)}{}^{i}{}_{j}$ is the expression (\ref{sigmaVdef}) evaluated on the leading order solution of the auxiliary fields and is given as:
\begin{align}
\Sigma_{a}^{(0)}{}^{i}{}_{j}=12 T^{(0)}_{jab}D_{c}T^{(0)bci}-\text{h.c}-\text{trace}
\end{align}
where $T^{(0)}$ is the leading order solution to the auxiliary T-field given in (\ref{LOaux}). In the same way the ($\mathcal{O}(\lambda)$) correction to the other auxiliary fields are given as:
\begin{align}
E_i^{(1)}&=0\;,\nn\\
A^{(1)}_{a}&=-2i\lambda\kappa^2T^{(0)}_{abi}D_{c}T^{(0)bci}+\text{h.c}\;, \nn\\
T_{ab}^{(1)i}&=9\lambda\kappa^4\mathcal{M}_{abj}^{(0)+}\xi^{(0)jK}\xi^{(0)iL}\eta_{KL}\;, \nn\\
D^{(1)i}{}_{j}&=18\lambda\kappa^2\left[T^{(0)i}\cdot \mathcal{M}_{j}^{(0)}\right]_{\textbf{8}}\;,
\end{align}
where, $\mathcal{M}_{abi}^{(0)}$ is the expression (\ref{MabDef}) evaluated on the leading order solutions of the auxiliary fields and is given as:
\begin{align}
\mathcal{M}_{abi}^{(0)}&=32D_{a}D^{c}T^{(0)}_{bci}+T_{ab}^{(0)j}T^{(0)}_{i}\cdot T^{(0)}_{j}
\end{align}
It is clear that if the leading order Maxwell's equation is satisfied, then $D_{c}T^{(0)bci}=0$. One can also show that if the leading order Einstein-Maxwell's equations (\ref{Ein_Max_eom},\ref{Max1}) are satisfied, then $\cM_{abi}^{(0)+}=0$ (refer to appendix-\ref{eqtrunc}) and hence all the above first order corrections and as a consequence all order corrections to the auxiliary field vanishes and the leading order solutions to the auxiliary fields (\ref{LOaux},\ref{L_eom}) become exact. As a result of which the truncation of the Lagrangian at the fourth order in derivatives (\ref{eq:higherderivativeaction}) is a consistent truncation. However, in order to arrive at this conclusion, one must show that the corrections to the dynamical Einstein-Maxwell's equation coming from the fourth order piece vanish when the leading order Einstein-Maxwell's equations (\ref{Ein_Max_eom},\ref{Max1}) are satisfied. In the context of Maxwell's equations, it is important to note that the corrections to the Maxwell's equation come via the corrections to the auxiliary T-field. In order to see this, consider the fields $\mathcal{A}_{\mu}^{I}$ and $T^{abi}$ in the Lagrangian (\ref{LVCSG}) as independent fields. The Maxwell's equation is obtained by varying the action w.r.t $\mathcal{A}_{\mu}^{I}$ which comes only from $\mathcal{L}_{V}$ and is given as:
\begin{align}
\nabla_{\mu}\left(\widehat{F}^{\mu\nu I}-T^{\mu\nu i}\xi_{i}^{I}-T^{\mu\nu}_{i}\xi^{iI}\right)=0
\end{align}
Inserting the first order corrections to $T_{\mu\nu}$ in the above equation, we obtain the corrections to the Maxwell's equation as:
\begin{align}
\nabla_{\mu}\widehat{F}^{\mu\nu I}&=3\lambda\kappa^2\nabla_{\mu}\left(\mathcal{M}^{(0)+}_{j}{}^{\mu\nu}\xi^{jI}+\text{h.c}\right)
\end{align}
Since $\mathcal{M}^{(0)+}_{abj}$ vanishes upon using the leading order Einstein-Maxwell's equation, it also implies that the corrections to the Maxwell's equation also vanishes. The corrected Einstein equations can be obtained from \eqref{eq:higherderivativeaction} and is given as
\begin{equation}\label{L_Ein}
	\begin{aligned}
		R_{\mu \nu}-\frac{1}{2}g_{\mu \nu}R&=\frac{3\kappa^2}{4}\left(2 g^{\rho \sigma}\hat{F}_{\mu \rho}\hat{F}_{\nu \sigma}-\frac{1}{2}g_{\mu \nu}\hat{F}\cdot \hat{F} \right)-24 \lambda\kappa^2 \left(2 \nabla^{\rho}\nabla^{\sigma} C_{\mu \rho \nu \sigma}-\left(R^{\rho \sigma}-\frac{1}{2}g^{\rho \sigma}R\right)C_{\mu \rho \nu \sigma}\right)\\
		&+144 \lambda\kappa^4 \left(-\frac{1}{2}g_{\mu \nu}R_{\rho \sigma} \hat{F}^{+I \rho}\;_{ \alpha}\hat{F}^{-J \alpha \sigma }\eta_{IJ} {-}\frac{1}{2}g_{\mu \nu}\nabla_{\alpha}\nabla_{\beta}(\hat{F}^{+I \alpha}\;_{ \rho}\hat{F}^{-J \rho \beta })\eta_{IJ} \right.\\
		&{-}\frac{1}{2}\nabla^2 (\hat{F}^{+I}_{\mu \rho}\hat{F}^{-J\rho}\;_{\nu})\eta_{IJ} {+ }\nabla_{\alpha}\nabla_{(\mu}(\hat{F}^{+I }\;_{\n) \rho}\hat{F}^{-J\rho \a})\eta_{IJ}+2 R_{\s (\mu }\hat{F}^{+I}_{\nu)}\;^{ \rho}\hat{F}_\r^{-J  \; \sigma}\eta_{IJ}\\
		&\left.+R_{ \alpha \beta}\hat{F}^{+ I \a}\;_{ (\m}\hat{F}_{\n)}^{-J \;\b}\eta_{IJ}\right)+54\lambda\kappa^6 \left({\frac{1}{2}} g_{\mu \nu}(\hat{F}^{+I}\cdot \hat{F}^{+K})(\hat{F}^{-J}\cdot \hat{F}^{-L})\eta_{KL}\eta_{IJ} \right)\\ 
		&+....
	\end{aligned}
\end{equation}
where (....) represents all the terms that trivially vanishes on imposing the leading order Maxwell's equations (\ref{Max1}). After some tedious but straightforward calculations, one can show that the $\mathcal{O}(\lambda)$ corrections to the Einstein's equation vanishes upon using the leading order Einstein-Maxwell's equation (\ref{Ein_Max_eom}, \ref{Max1}) (Please refer to appendix-\ref{eqtrunc}). 

As a result of the above discussions, we can conclude that for the case of pure $\cN=3$ supergravity, we can consistently truncate our action at the fourth order in derivatives where the auxiliary fields as well as the dynamical fields (graviton and graviphoton) are given by their leading order two derivative equations of motion. Such a feature has been observed in the case of minimal $\cN=2$ ungauged \cite{Charles:2016wjs,Charles:2017dbr} as well as gauged supergravity \cite{Bobev:2021oku}. However, when one considers matter coupled supergravity theories, one cannot consistently truncate at the fourth order in derivatives, and one indeed needs to consider the full derivative expansion of the action.
\section{Conclusion and future directions}
Supergravity theories often arise as a low energy limit of string/M-theory and are useful for understanding the physics of black holes in these theories. Therefore, the classification of supergravity theories is of interest. The superconformal approach has been crucial in the construction and classification of various supergravity invariants in four, five, and six dimensions. In this paper, we have used the superconformal approach to derive pure $\mathcal{N}=3$ supergravity in four dimensions along with four derivative corrections. To do this, we used the $\mathcal{N}=3$ Weyl multiplet constructed in \cite{vanMuiden:2017qsh, Hegde:2018mxv} and the $\mathcal{N}=3$ conformal supergravity action constructed in \cite{Hegde:2021rte}. To compensate for additional symmetries in the superconformal theory, we constructed $\mathcal{N}=3$ vector multiplet coupled to conformal supergravity by using supersymmetric reduction of $\mathcal{N}=4$ vector multiplet coupled to conformal supergravity. We constructed the action for the $\mathcal{N}=3$ vector multiplet coupled to conformal supergravity by using the density formula constructed in \cite{Hegde:2021rte} using the covariant superform method and by adding terms proportional to the vector multiplet equations of motion which we obtained by performing supersymmetric reduction of $\cN=4$ vector multiplet equations of motion coupled to conformal supergravity. We then used the gauge fixing procedure to obtain pure quadratic as well as higher derivative supergravity by using the actions for the Weyl and the vector multiplet. We showed that while eliminating auxiliary fields and obtaining the action as an expansion in derivatives, one can consistently truncate the action to fourth order in derivatives, where the auxiliary fields as well as the dynamical fields such as the graviton and the graviphoton obeys their leading order equations of motion. The main result of this paper is the $\mathcal{N}=3$ Poincar\'e supergravity action up to fourth order in derivatives given in \eqref{eq:higherderivativeaction}. And some of our important intermediate results are: $\cN=3$ vector multiplet coupled to conformal supergravity and its complete supersymmetry transformation (\ref{N3susy}), its equations of motion (\ref{ferm_eom}, \ref{scalareom}) and its action (\ref{lag}). 

While we have constructed the pure supergravity action by using three compensating vector multiplets, our action for vector multiplets coupled to conformal supergravity holds for arbitrary number of vector multiplets. One can therefore consider the action for $3+n_v$ vector multiplets coupled to conformal supergravity to produce matter coupled $\cN=3$ Poincar\'e supergravity. When the action for $3+n_v$ vector multiplets is considered along with the pure $\mathcal{N}=3$ conformal supergravity action, we can obtain higher derivative corrections to the two derivative results in \cite{Castellani:1985ka}. Further, minimal gauged supergravity theories have been recently shown to produce surprising insights into holography \cite{Bobev:2020egg,Bobev:2021oku}. It will be interesting to construct minimal gauged $\mathcal{N}=3$ supergravity and obtain these insights for $\mathcal{N}=3$ theories.

For conformal supergravities with eight supercharges in five \cite{Bergshoeff:2001hc} and six dimensions \cite{Bergshoeff:1985mz}, it was found long ago that there exist two versions of the Weyl multiplet. While they both contain all the gauge fields for the superconformal algebra, their auxiliary field content differs. In particular, one of them contains a scalar field of Weyl weight $+1$ and its superpartner, which can be used to internally gauge fix dilatation and S-supersymmetry without resorting to any compensating multiplets. Working with this multiplet reduces the number of compensating multiplets required for going from conformal supergravity to Poincar{\'e} supergravity. This was dubbed the dilaton Weyl multiplet, while the other as the standard Weyl multiplet. Dilaton Weyl multiplet was also constructed for four dimensional $\mathcal{N}=2$ conformal supergravity in \cite{Butter:2017pbp}. Dilaton Weyl multiplets have the advantage that it allows us to directly construct the supersymmetrization of a Riemann square term which can be then used to construct the supersymmetrization of arbitrary curvature squared invariants. This was used to construct supersymmetrization of arbitrary curvature squared invariants in $\cN=1$ supergravity in five dimensions \cite{Ozkan:2013nwa} as well as $\cN=2$ supergravity in four dimensions  \cite{Mishra:2020jlc} using the respective dilaton Weyl multiplets. It is an interesting open question if the dilaton Weyl multiplet exists for $\mathcal{N}=3$ and $\mathcal{N}=4$ conformal supergravity theories in four dimensions. In recent work (which is to appear soon), where one of the authors is involved \cite{Ciceri2022}, it was found that the dilaton Weyl multiplet in $\mathcal{N}=4$ conformal supergravity manifests a $USp(4)$ R-symmetry. One can imagine that if one performs a supersymmetric truncation of this Weyl multiplet, one could perhaps be able to construct a dilaton Weyl multiplet for $\cN=3$ conformal supergravity and hence the R-symmetry that would be manifest in an $\cN=3$ dilaton Weyl multiplet should be some subgroup of $USp(4)$.
On the other hand, if one constructs directly in $\cN=3$ using the standard method of coupling a vector multiplet to the standard Weyl multiplet and using the equations of motion of the vector multiplet to eliminate the auxiliary fields of the standard Weyl multiplet in terms of the auxiliary fields of the dilaton Weyl multiplet, one would expect that the R-symmetry that would be manifest in $\cN=3$ dilaton Weyl multiplet would be a subgroup of $SU(3)$. Hence one might think that some subgroup that would lie in the overlap of $USp(4)$ and $SU(3)$ would be the R-symmetry manifested in $\cN=3$ dilaton Weyl multiplet. It would be interesting to investigate this dilaton Weyl multiplet in $\mathcal{N}=3$ conformal supergravity and use it for an alternate formulation of $\cN=3$ Poincar{\'e} supergravity and find its relation to the Poincar{\'e} supergravity constructed in this paper. We leave these questions for future work.

\section*{Acknowledgements}
We would like to thank Dileep Jatkar for useful discussion. SH thanks IIT Ropar, IISER Mohali, IISER Thiruvananthapuram and AEI Potsdam for hospitality during the course of this work. This work is partially supported by SERB core research grant grant CRG/2018/002373, Government of India.

\appendix
\section{Conventions}\label{appendix-conventions}

In this paper, curved indices are denoted by $\m, \n = 0, 1,..$ while local Lorentz indices are denoted by $a, b, = 0, 1,..$ respectively. We always (anti)symmetrize with weight one, for example:
\bea
T_{[ab]}	= \frac{1}{2} \left( T_{ab} - T_{ba}\right), \quad T_{[ab]}	= \frac{1}{2} \left( T_{ab} - T_{ba}\right)
\eea
The completely antisymmetric tensor satisfies:
\bea
\varepsilon^{abcd}= e^{-1} \varepsilon^{\m \n \r \s} e_\m^a e_\n^b e_\r^c e_\s^d , \quad \varepsilon^{0123} = i 
\eea
which implies $\gamma_{ab} = -\frac{1}{2} \varepsilon_{abcd} \gamma^{cd} \gamma_{5}$. 
The dual of an antisymmetric tensor field $F_{ab}$ is given by:
\bea
\tilde{F}_{ab} = \frac{1}{2} \varepsilon_{abcd} F^{cd}
\eea

and the (anti)selfdual part of $F_{ab}$ reads:
\be
F^{\pm}_{ab} = \frac{1}{2} \left( F_{ab} \pm \tilde{F}_{ab} \right)
\ee
Our convention of Riemann tensor in terms of the spin-connection is given as:
\begin{align}\label{Riem}
R_{\mu\nu}{}^{ab}=2\partial_{[\mu}\omega_{\nu]}^{ab}-2\omega_{[\mu}^{ac}\omega_{\nu]c}{}^{b}
\end{align}
Recall that we had defined a covariant derivative which is covariant w.r.t general coordinate transformation along with local Lorentz transformations in (\ref{covgct}). Under this covariant derivative, $e_{\mu}^{a}$ and subsequently the metric $g_{\mu\nu}$ is covariantly constant. Hence the Christoffel connection defined in (\ref{christoff}) is given as:
\begin{align}
\Gamma^\r_{\m \n} &= \frac{1}{2} g^{\r \s} \left( \partial_\m g_{\s \n} + \partial_\n g_{\m \s} - \partial_\s g_{\m \n} \right)
\end{align}
Further using $[\nabla_{\mu},\nabla_{\nu}]e_{\rho}^{a}=0$, we get the Riemann tensor defined in (\ref{Riem}) completely written in terms of curved indices as:
\begin{align}
R^\s{}_{\m \n \r}\equiv R_{\n \r}{}^{ab}e^{\s}_{a}e_{\m b}=\partial_\r \Gamma^\s_{\m \n} - \partial_\n \Gamma^\s_{\m \r} + \Gamma^\s_{\lambda \r} \Gamma^\lambda_{\m \n} - \Gamma^\s_{\n \lambda} \Gamma^\lambda_{\m \r}
\end{align}
The Ricci tensor and Ricci scalar are defined as:
\begin{align}
R_{\m \n} =R_{\mu\rho}{}^{ab}e^{\rho}_{b}e_{\nu a}= R^\r{}_{\m \r \n}, \quad R = g^{\m \n} R_{\m \n}
\end{align}
The above convention of Riemann tensor is consistent with the negative sign appearing with the Einstein-Hilbert term in (\ref{vPLag}). That is the reason why we chose the compensating vector multiplets with the wrong sign of the kinetic term. The Weyl tensor is the trace-free part of the Riemann tensor and is given as,
\be\label{Weyl}
C_{\m \n}{}^{ab} = R_{\m \n}{}^{ab} - 2 \delta^{[a}_{[\m} R^{b]}_{\n]} + \frac{1}{3} R  \delta^{[a}_{[\m} \delta^{b]}_{\n]} 
\ee

\section{Transformations of the dependent gauge fields and Bianchi identity}\label{trans_dep}

For the purpose of completeness, we provide explicit expressions of the fully supercovariant curvatures in $\cN=3$ conformal supergravity according to the conventions followed in \cite{Hegde:2021rte}:
\bea
R(P)_{\m \n}{}^a &=& 2 \partial_{[\m}e_{\n]}{}^a - 2 \omega_{[\m}{}^{ab} e_{\n]b} + 2 b_{[\m}e_{\n]}{}^a -\frac12 \left(\bar{\psi}_{\m}^i \gamma^a \psi_{\n i} + h.c. \right)\,\nn\\ 
{R}(Q)_{\mu\nu}{}^i &=&2\left(\partial_{[\mu}- \frac{1}{4}\omega _{[\mu}{}^{ab}\gamma _{ab}+\frac12 b_{[\mu} +\ft12\rmi{A}_{[\mu} \right)\psi _{\nu ]}^i - 2  V_{[\mu}{}^i{}_j{}\psi _{\nu ]}{}^{j}\nn \\
&-& \frac18 \varepsilon^{ijk} \gamma \cdot T_{ j} \gamma _{[\mu} \psi _{\nu ]\,k} + \frac{1}{2}\varepsilon^{ijk} {\Lambda}_L \bar{\psi}_{\mu \,j} \psi_{\nu\, k}{}
- \gamma_{[\mu}{}\phi _{\nu ]}^{i}\,,\nn\\
{R}(M)_{\m\n}{}^{ab} &=& 2\partial_{[\mu}\omega _{\nu ]}{}^{ab}- 2\omega _{[\mu }{}^{ac}\omega _{\nu ]c}{}^b-4 f_{[\m}{}^{[a}e_{\n]}{}^{b]}  \\
&+&\left(\frac12 \bar \psi _{[\mu} ^i \gamma ^{ab}\phi _{\nu] i}-\frac14 \varepsilon _{ijk}\bar \psi _\mu  ^i\psi _\nu ^jT^{ab\,k} -\bar \psi _{[\mu }^i\gamma _{\nu]} \widehat{R}^{ab}(Q)_i+ h.c.\right)\,,\nn \\
{R}( V)_{\mu\nu}{}^i{}_j &=& 2\partial_{[\mu } V_{\nu ]}{}^i{}_j +  { V}_{[\mu}{}^i{}_k {}{V}_{\nu ]}{}^k{}_j\nn \\
&+&\left(-\bar \psi _{[\mu }^{i} \phi_{\nu] j} + \frac{1}{48} \bar{\psi}_{[\mu }^{i}\gamma_{\nu]} \zeta_{j}- \frac{1}{16} \varepsilon_{jkl}{} \bar{\psi}_{[\mu }^{k} \gamma_{\nu]}{} \chi^{i l}  \right.\nn\\
&&\left. - \frac{1}{16} \varepsilon_{k l j}{} E^i \bar{\psi}_\mu^{k} \psi_{\nu}{}^l + \frac{1}{16}\bar{\psi}_{[\mu}^i\gamma \cdot T_j \gamma_{\nu]} \Lambda_R -\frac{1}{16}\bar{\psi}_{[\mu}{}^i\gamma_{\nu]}{} \Lambda_R E_j\right.\nn\\
&&\left. - \frac{1}{8} \bar{\psi} _{[\mu}^{i} \gamma^{a} \psi_{\nu] j}{} \bar{\Lambda}_L\gamma_{a} \Lambda_R    -\, h.c. - \text{trace}\right)\, \nn \\
R(A)_{\m \n} &=& 2\partial_{[\mu } A_{\nu ]} -i \left(\frac16 \bar \psi _{[\mu }^{i} \phi_{\nu] j} + \frac{1}{36} \bar{\psi}_{[\mu }^{i}\gamma_{\nu]} \zeta_{i}  + \frac{1}{24} \varepsilon_{k l j}{} E^j \bar{\psi}_\mu^{k} \psi_{\nu}{}^l\right.\nn\\
&&\left. + \frac{1}{12}\bar{\psi}_{[\mu}^i\gamma \cdot T_i \gamma_{\nu]} \Lambda_R +\frac{1}{12}\bar{\psi}_{[\mu}^i\gamma_{\nu]}{} \Lambda_R E_i - \frac{1}{6} \bar{\psi} _{[\mu}^{i} \gamma^{a} \psi_{\nu] i}{} \bar{\Lambda}_L\gamma_{a} \Lambda_R + h.c. \right)
\eea
As per the conventions followed in \cite{Hegde:2021rte}, the set of curvature constraints required to reduce the superconformal gauge multiplet to the independent fields listed in Table-\ref{Table-Weyl} are given as:
\bea
R(P)_{\m\n}{}^a = 0\,\nn \\
R(M)_{\m\n}{}^{ab} e^\n{}_b = 0\,\nn \\
\gamma^\m R(Q)_{\m\n}{}^i = 0
\eea
 The expressions for dependent gauge fields $\omega_\mu^{ab}$, $ f_\mu{}^a$ and $\phi_\mu^i$ are obtained by solving the above constraints. Their complete Q and S-supersymmetry transformations is given as follows:
\begin{align}\label{deptransf}
\delta \omega_\mu^{ab}&=-\frac{1}{2}\bar{\epsilon}^i\gamma^{ab}\phi_{\mu i}+\frac{1}{2}\varepsilon_{ijk}\bar{\epsilon}^i\psi_\mu^jT^{ab k}+\bar{\epsilon}^i\gamma_\mu R(Q)^{ab}{}_i-\frac{1}{2}\bar{\eta}^i\gamma^{ab}\psi_{\mu i}+\thc\nonumber\\
\delta \phi_\mu^i&= -\frac{i}{12}(\gamma_\mu\gamma\cdot R(A)-3\gamma\cdot R(A)\gamma_\mu)\epsilon^i-\frac{1}{6}(3\gamma\cdot R(V){}^i{}_j\gamma_\mu-\gamma_\mu\gamma\cdot R(V){}^i{}_j)\epsilon^j\nonumber\\
&\quad-\frac{1}{4}\varepsilon^{ijk}\bar{\Lambda}_L\gamma_{\mu}R_{ab}(Q)_k\gamma^{ab}\epsilon_j+ \frac{1}{32}\gamma\cdot T^{[i}\gamma_{\mu}\gamma\cdot T_j\epsilon^{j]}+\frac{1}{24}\varepsilon^{ijk}(\gamma_{\mu}\gamma\cdot\slashed{D}T_j-3\slashed{D}\gamma\cdot T_j\gamma_\mu)\epsilon_k\nonumber\\
& \quad -\frac{1}{12}\bar{\epsilon}^{[i}\psi_{\mu}^{k]}\zeta_{k}+\frac{1}{32}\Big(\bar{\epsilon}^{i}\gamma_{a}\psi_{\mu j}-\delta^{i}_{j}\bar{\epsilon}^{k}\gamma_{a}\psi_{\mu k}+\bar{\epsilon}_{j}\gamma_{a}\psi_{\mu}^{i}-\delta^{i}_{j}\bar{\epsilon}_{k}\gamma_{a}\psi_{\mu}^{k}\Big)\gamma^{a}\Lambda_{L}E^{j} \nonumber \\
&\quad +\frac{1}{96}\Big(\bar{\epsilon}^{i}\gamma_{a}\psi_{\mu j}-\delta^{i}_{j}\bar{\epsilon}^{k}\gamma_{a}\psi_{\mu k}+\bar{\epsilon}_{j}\gamma_{a}\psi_{\mu}^{i}-\delta^{i}_{j}\bar{\epsilon}_{k}\gamma_{a}\psi_{\mu}^{k}\Big)\gamma^{a}\zeta^{j}-\frac{1}{4}\bar{\epsilon}^{[i}\psi_{\mu}^{j]}E_{j}\Lambda_R\nonumber \\
&\quad  -\frac{1}{32}\varepsilon^{ijk}\Big(\bar{\epsilon}^{l}\gamma_{a}\psi_{\mu k}+\bar{\epsilon}_{k}\gamma_{a}\psi_{\mu}^{l}\Big)\gamma^{a}\chi_{jl}-\frac{1}{8}\varepsilon_{jkl}\bar{\epsilon}^{j}\psi_{\mu}^{k}\chi^{il}-\frac{1}{2}\varepsilon^{ijk}\bar{\epsilon}_{j}\psi_{\mu k}\slashed{D}\Lambda_{L}\nonumber \\
&\quad -\frac{1}{16}\Big(\bar{\epsilon}^{i}\gamma_{a}\psi_{\mu j}-\delta^{i}_{j}\bar{\epsilon}^{k}\gamma_{a}\psi_{\mu k}+\bar{\epsilon}_{j}\gamma_{a}\psi_{\mu}^{i}-\delta^{i}_{j}\bar{\epsilon}_{k}\gamma_{a}\psi_{\mu}^{k}\Big)\gamma\cdot T^{j}\gamma^{a}\Lambda_{L}-\frac{1}{4}\varepsilon^{ijk}\bar{\epsilon}_{j}\gamma_{a}\phi_{\mu k}\gamma^{a}\Lambda_{L}\nonumber \\
&\quad+2\mathcal{D}_{\mu}\eta^{i}-\frac{1}{24}\varepsilon^{ijk}\gamma_\mu\gamma\cdot T_j\eta_k+\frac{1}{4}\varepsilon^{ijk}\bar{\eta}_{j}\gamma_{a}\psi_{\mu k}\gamma^{a}\Lambda_{L}\nonumber \\
\delta f_\mu{}^a&=-\bar{\epsilon}^i\gamma_\mu D_b R(Q)^{ab}{}_i+\frac{1}{4}\bar{\epsilon}^i {\tilde{R}(S)}^a{}_{\mu i}+\frac{1}{2}\varepsilon_{ijk}T_{\mu b}{}^k\bar{\epsilon}^iR(Q)^{abj}-\frac{i}{6}\bar{\epsilon}^{k}\gamma_{b}\psi_{\mu k}\tilde{R}(A)^{ab}\nonumber \\
& \quad +\frac{1}{3}\bar{\epsilon}^{i}\gamma_{b}\psi_{\mu j}\tilde{R}(V)^{ab}{}^{j}{}_{i}+\frac{1}{64}\bar{\epsilon}^{[i}\gamma\cdot T_{i}\gamma^{a}\gamma\cdot T^{j]}\psi_{\mu j}-\frac{1}{3}\varepsilon_{ijk}\bar{\epsilon}^{i}\psi_{\mu}^{j}D_{b}T^{ab k}+\frac{1}{48}\varepsilon_{ijk}\bar{\epsilon}^{i}\gamma^{a}\gamma\cdot T^{j}\phi_{\mu}^{k}\nonumber \\
&\quad +\frac{1}{2}\bar{\eta}^{i}\gamma^{a}\phi_{\mu i} -\frac{1}{4}\bar{\eta}^{i}R(Q)_{\mu}{}^{a}{}_{i}-\frac{1}{48}\varepsilon_{ijk}\bar{\eta}^{i}\gamma\cdot T^{k}\gamma^{a}\psi_{\mu}^{j}+\thc,\nonumber\\
\end{align}
{\tiny }The above mentioned supersymmetry transformations of the dependent gauge fields can be obtained from the independent ones as explained in \cite{Freedman:2012zz}. The Bianchi identities satisfied by the curvatures as shown below: 
\begin{align}\label{Bianchi}
R(D)_{ab}&=0 \nonumber \\
R(M)_{abcd}&=R(M)_{cdab}\nonumber\\
\varepsilon^{aecd}R(M)_{cdeb}&=0\nonumber \\
\frac{1}{4}\varepsilon^{abcd}\varepsilon^{efgh}R(M)_{abef}&=R(M)^{cdgh}\nonumber\\
\varepsilon^{cdef}D_{b}D_{d}R(M)_{efab}&=0 \nonumber \\
R(K)_{ab}{}^{c}&=-D_{e}R(M)_{ab}{}^{ec}\nonumber \\
\varepsilon^{abcd}D_{b}R(V)_{cd}{}_{j}{}^{i}{}&=\frac{1}{16}\bar{\Lambda}_{L}\gamma_{b}\gamma\cdot T^{i}R(Q)^{ab}_{j}+\text{(h.c; traceless)}\nonumber \\
\varepsilon^{abcd}D_{b}R(A)_{cd}&=-\frac{i}{12}\bar{\Lambda}_{L}\gamma_{b}\gamma\cdot T^{j}R(Q)^{ab}{}_{j}-\frac{i}{12}\bar{\Lambda}_{R}\gamma_{b}\gamma\cdot T_{j}R(Q)^{ab}{}^{j}\nonumber \\
D_{a}R(Q)^{abi}&=-\frac{1}{4}\varepsilon^{abcd}\gamma_{a}R(S)_{cd}^{i}\nonumber \\
R(Q)_{ab}^{+i}&=0 \nonumber \\
R(S)_{ab}^{-i}&=\slashed{D}R(Q)_{ab}^{i}\nonumber \\
\gamma^{ab}R(S)_{ab}^{i}&=0\nonumber \\
\gamma^{a}R(S)_{ab}^{+i}&=0 \nonumber \\
\varepsilon^{abcd}D_{b}R(S)_{cd}^{i}&=\frac{1}{12}\varepsilon^{ijk}\gamma^{a}T_{k}\cdot R(S)_{j}+\frac{1}{3}\varepsilon^{ijk}T^{ab}_{k}D^{d}R(Q)_{dbj}\nonumber \\
&\quad -\frac{1}{3}\gamma^{a}R(V)^{i}{}_{j}\cdot R(Q)^{j}-\frac{i}{6}\gamma^{a}R(A)\cdot R(Q)^{i} \nonumber \\
&\quad -\frac{1}{3}\varepsilon^{ijk}D^{g}T_{gck}R(Q)^{ac}_{j}+\frac{1}{32}\gamma\cdot T^{[l}\gamma^{a}T_{l}\cdot R(Q)^{i]}
\end{align}

\section{Truncation of Equations of motion}\label{eqtrunc}
In this appendix we show that the $\mathcal{O}(\lambda)$ corrections to the $T_{ab}^i$ and Einstein's equation vanishes upon using the leading order solution $T_{ab}^{(0) i}$ given in (\ref{LOaux}) and Einstein-Maxwell's equation (\ref{Ein_Max_eom}, \ref{Max1}). First we will consider $\mathcal{O}(\lambda)$ corrections to the $T_{ab}^i$ given as:

\be\label{LT}
T_{ab}^{(1)i}=9\lambda\kappa^4\mathcal{M}_{abj}^{(0)+}\xi^{(0)jK}\xi^{(0)iL}\eta_{KL}\;,
\ee
where, $\mathcal{M}_{abi}^{(0)}$ is given as:
\begin{align}\label{Mo}
\mathcal{M}_{abi}^{(0)}&=32D_{a}D^{c}T^{(0)}_{bci}+T_{ab}^{(0)j}T^{(0)}_{i}\cdot T^{(0)}_{j}
\nn \\
&=32\cD_{a}\cD^{c}T^{(0)}_{bci} -32f_a{}^c T^{(0)}_{bci}  +T_{ab}^{(0)j}T^{(0)}_{i}\cdot T^{(0)}_{j}
\end{align}
where covariant derivative $D_{a}$ appearing in the first line is the fully supercovariant derivative. Since we are interested in only the bosonic part, we can replace it with $D_{a}=\mathcal{D}_{a}+\text{K-covariantization}$ in the second line where $\mathcal{D}_{a}$ is covariant w.r.t all the standard gauge transformations (dilatation, local Lorentz transformation, and R-symmetry). The K-gauge field  $f_{\mu}^{a}$ in terms of the Ricci tensor and Ricci scalar is given as:
\begin{align}
f_{\mu}^{a}=\frac{1}{2}R_{\mu}{}^{a}-\frac{1}{12}Re_{\mu}^{a}
\end{align}
Upon using Einstein-Maxwell's equation (\ref{Ein_Max_eom}), one can write Ricci tensor and Ricci scalar as shown below:
\bea\label{Rmn}
R_{\m \n} = - 3 \kappa^2 \widehat{F}^{+I}_\m{}^\a \widehat{F}^{-J}_{\a \n} \eta_{I J}, \quad R=0 
\eea
Upon substituting $T_{ab}^{(0) i}$ and the expression Ricci tensor and Ricci scalar from above in (\ref{Mo})  we get,
\begin{align}
\mathcal{M}_{abi}^{(0)}&= - 192\; \xi^I_i \cD_{a}\cD^{c} \widehat{F}_{bc}^{-J} \eta_{IJ} + 288\;\xi^I_i \widehat{F}_{c \a}^{+K} \widehat{F}^{-L[c}{}_{[a} \widehat{F}^{-J}_{b]}{}^{\a]} \eta_{I J} \eta_{KL} \nn \\
&\quad + 72\;\xi^I_i \widehat{F}_{ab}^{+K} \left( \widehat{F}^{-L} \cdot \widehat{F}^{-J} \right) \eta_{IJ} \eta_{KL}
\end{align}
To arrive at the last term we use Eq (\ref{M_minimal}) and the D-gauge condition given in (\ref{Dgauge}).
We notice that the first term of the above equation vanishes upon using the Maxwell's equation together with the Bianchi identity given in (\ref{Max}). The second term however, can further be simplified on the using folowing identity:
\be
G^{\pm}_{[a[c} H^{\pm}_{d]b]} = \pm \frac{1}{8} G^{\pm}_{ef} H^{\pm ef} \varepsilon_{abcd} - \frac{1}{4} \left( G^\pm_{ab} H^\pm_{cd} + G^\pm_{cd} H^\pm_{ab}\right)
\ee
We have,
\begin{align}
\mathcal{M}_{abi}^{(0)}&=  - 36\;\xi^I_i \widehat{F}_{c \a}^{+K} \left(\widehat{F}^{-L} \cdot \widehat{F}^{-J} \varepsilon^{c\a}{}_{ab} \right)\eta_{I J} \eta_{KL} - 144 \; \xi^I_i \widehat{F}_{c \a}^{+K} \left(\widehat{F}^{-L}_{ab} \cdot \widehat{F}^{-J c\a} \right)\eta_{I J} \eta_{KL} \nn \\
&\quad + 72\;\xi^I_i \widehat{F}_{ab}^{+K} \left( \widehat{F}^{-L} \cdot \widehat{F}^{-J} \right) \eta_{IJ} \eta_{KL}
\end{align}
Now, for the $\mathcal{O}(\lambda)$ corrections to the $T_{ab}^i$ given in (\ref{LT}), we need to take selfdual projection of $\mathcal{M}_{abi}^{(0)}$, which yields
\bea
\mathcal{M}_{abi}^{(0)+} &=  - 36\;\xi^I_i \widehat{F}_{c \a}^{+K} \left(\widehat{F}^{-L} \cdot \widehat{F}^{-J} \right) \left(\varepsilon^{c\a}{}_{ab} \right)^+ \widehat{F}_{c \a}^{+K} \eta_{I J} \eta_{KL}
\nn \\
&\quad + 72\;\xi^I_i \widehat{F}_{ab}^{+K} \left( \widehat{F}^{-L} \cdot \widehat{F}^{-J} \right) \eta_{IJ} \eta_{KL} = 0 
\eea
where we have used the fact that $\left(\varepsilon^{c\a}{}_{ab} \right)^+ \widehat{F}_{c \a}^{+K} = 2 \widehat{F}_{a b}^{+K}$. Thus, we showed 
\be\label{LT}
T_{ab}^{(1)i}=9\lambda\kappa^4\mathcal{M}_{abj}^{(0)+}\xi^{(0)jK}\xi^{(0)iL}\eta_{KL} = 0
\ee

Similarly, we will show now that the $\mathcal{O}(\lambda)$ corrections to the Einstein's equation (\ref{L_Ein}) also vanishes. We consider, the non-trivial part of the corrected Einstein's equation as shown below:

\begin{equation}\label{L_Ein}
\begin{aligned}
E_c^{(1)} &= -24 \lambda\kappa^2 \left(2 \nabla^{\rho}\nabla^{\sigma} C_{\mu \rho \nu \sigma}-\left(R^{\rho \sigma}-\frac{1}{2}g^{\rho \sigma}R\right)C_{\mu \rho \nu \sigma}\right)\\
&+144 \lambda\kappa^4 \left(-\frac{1}{2}g_{\mu \nu}R_{\rho \sigma} \hat{F}^{+I \rho}\;_{ \alpha}\hat{F}^{-J \alpha \sigma }\eta_{IJ} {-}\frac{1}{2}g_{\mu \nu}\nabla_{\alpha}\nabla_{\beta}(\hat{F}^{+I \alpha}\;_{ \rho}\hat{F}^{-J \rho \beta })\eta_{IJ} \right.\\
&{-}\frac{1}{2}\nabla^2 (\hat{F}^{+I}_{\mu \rho}\hat{F}^{-J\rho}\;_{\nu})\eta_{IJ} {+ }\nabla_{\alpha}\nabla_{(\mu}(\hat{F}^{+I }\;_{\n) \rho}\hat{F}^{-J\rho \a})\eta_{IJ}+2 R_{\s (\mu }\hat{F}^{+I}_{\nu)}\;^{ \rho}\hat{F}_\r^{-J  \; \sigma}\eta_{IJ}\\
&\left.+R_{ \alpha \beta}\hat{F}^{+ I \a}\;_{ (\m}\hat{F}_{\n)}^{-J \;\b}\eta_{IJ}\right)+54\lambda\kappa^6 \left({\frac{1}{2}} g_{\mu \nu}(\hat{F}^{+I}\cdot \hat{F}^{+K})(\hat{F}^{-J}\cdot \hat{F}^{-L})\eta_{KL}\eta_{IJ} \right)
\end{aligned}
\end{equation}
Upon writing the Weyl tensor in terms of Riemann tensor, Ricci tensor and Ricci scalar as given in (\ref{Weyl}), the first two terms of (\ref{L_Ein}) takes the following form
\bea\label{nc}
- 48 \lambda\kappa^2 \nabla^\r \nabla^\s C_{\m \r \n \s} &=&  -24\lambda\kappa^2 \left( \nabla^2 R_{\m \n} - g_{\m \n} \nabla^\r \nabla^\s R_{\r \s} - R_{\m \r \n \s} R^{\r \s} + R_{\s(\m}R_{\n)}{}^\s  \right)\nn \\ 
24 \lambda\kappa^2 R^{\r \s}  C_{\m \r \n \s} &=& 12 \lambda\kappa^2 \left( 2  R_{\m \r \n \s} R^{\r \s} - g_{\m \n} R^{\r \s} R_{\r \s} + 2 R_{\s(\m}R_{\n)}{}^\s \right)
\eea
To arrive at first equation of (\ref{nc}), we have used the following relations:
\be
\nabla^{\s} R_{\m \r \n \s} = \nabla_{\r} R_{\m \n} - \nabla_{\m} R_{\r \n}, \;\;\;\quad \left[ \nabla_\n , \nabla_\m \right] R^{\a \r} =  R^\a{}_{\s \m\n} R^{ \s \r}+ R^\r{}_{\s \m\n} R^{ \a \s}
\ee 
On substituting back (\ref{nc}) in (\ref{L_Ein}) and further writing the expressions for Ricci tensor as given in (\ref{Rmn}), we get
\begin{equation}\label{e_c}
\begin{aligned}
E_c^{(1)} &= \cancel{72 \lambda\kappa^4  \nabla^{2} (\hat{F}^{+I}_{\mu \rho}\hat{F}^{-J\rho}\;_{\nu})\eta_{IJ}} -\cancelto{0}{ 72 \lambda\kappa^4 g_{\mu \nu}\nabla_{\alpha}\nabla_{\beta}(\hat{F}^{+I \alpha}\;_{ \rho}\hat{F}^{-J \rho \beta })\eta_{IJ}} -12 \lambda\kappa^2 g_{\m \n} R^{\r \s} R_{\r \s} \\
&-144 \lambda\kappa^4 R_{\m \r \n\s} \hat{F}^{+I \r \a} \hat{F}^{-J}_{\a}{}^\s \eta_{IJ}   -72 \lambda\kappa^4 g_{\mu \nu}R_{\rho \sigma} \hat{F}^{+I \rho}\;_{ \alpha}\hat{F}^{-J \alpha \sigma }\eta_{IJ} {-}72 \lambda\kappa^4 \cancelto{0}{ g_{\mu \nu}\nabla_{\alpha}\nabla_{\beta}(\hat{F}^{+I \alpha}\;_{ \rho}\hat{F}^{-J \rho \beta })\eta_{IJ}} 
\\
&\cancel{-72 \lambda\kappa^4 \nabla^2 (\hat{F}^{+I}_{\mu \rho}\hat{F}^{-J\rho}\;_{\nu})\eta_{IJ}} {+ 144  \lambda\kappa^4  }\nabla_{\alpha}\nabla_{(\mu}(\hat{F}^{+I }\;_{\n) \rho}\hat{F}^{-J\rho \a})\eta_{IJ} +288\lambda\kappa^4 R_{\s (\mu }\hat{F}^{+I}_{\nu)}\;^{ \rho}\hat{F}_\r^{-J  \; \sigma}\eta_{IJ}\\
&  + 144\lambda\kappa^4 R_{ \r \s}\hat{F}^{+ I \r}\;_{ (\m}\hat{F}_{\n)}^{-J \;\s}\eta_{IJ} +54\lambda\kappa^6 \left({\frac{1}{2}} g_{\mu \nu}(\hat{F}^{+I}\cdot \hat{F}^{+K})(\hat{F}^{-J}\cdot \hat{F}^{-L})\eta_{KL}\eta_{IJ} \right)
\end{aligned}
\end{equation}
The second and sixth term goes to zero as a consequence of (\ref{Max}).  We now focus on the second term in the third line of (\ref{e_c}). 
\bea\label{c13}
144 \nabla_{\alpha}\nabla_{(\mu}(\hat{F}^{+I }\;_{\n) \rho}\hat{F}^{-J\rho \a})\eta_{IJ}
&=& - 144 \left( \left[ \nabla_\a , \nabla_{(\m} \right] F^{+I \a \r} \hat{F}^{-J}_{\n) \r} + \hat{F}^{-J}_{\n \r} \left[ \nabla_\a , \nabla_{(\m} \right] F^{-J \a \r} \hat{F}^{+I}_{\n) \r} \right)\eta_{IJ} \nn \\
&=& -144 R_{\s (\mu }\hat{F}^{+I}_{\nu)}\;^{ \rho}\hat{F}_\r^{-J  \; \sigma}\eta_{IJ} 
+ 144 R_{\m \r \n \s} \hat{F}^{+I \r \a} \hat{F}^{-J}_\a{}^\s \eta_{I J}
\eea
The reason we could express the above term as commutators is because the term $(\hat{F}^{+I }\;_{\n \rho}\hat{F}^{-J\rho \a})$ is symmetric in $\a$ and $\n$ thus, the contribution coming from the term where $\nabla_{\alpha}$ acts on it vanishes as a consequence of (\ref{Max}).
Thereafter, for simplification, we will take the first term of the last line in (\ref{e_c}) and re-express it as
\be\label{c14}
144 R_{ \r \s}\hat{F}^{+ I \r}\;_{ (\m}\hat{F}_{\n)}^{-J \;\s}\eta_{IJ} = -144 R_{\s (\mu }\hat{F}^{+I}_{\nu)}\;^{ \rho}\hat{F}_\r^{-J  \; \sigma}\eta_{IJ} + 72 g_{\m \n} R_{\rho \sigma} \hat{F}^{+I \rho}{}_{ \alpha}\hat{F}^{-J \alpha \sigma }\eta_{IJ}
\ee
Now, upon substituting (\ref{c13}) and (\ref{c14}) in the equation (\ref{e_c}), we get 
\bea
E_c^{(1)} &=   -12 \lambda \kappa^2 g_{\m \n} R^{\r \s} R_{\r \s}   +27 \lambda \kappa^6 g_{\mu \nu}(\hat{F}^{+I}\cdot \hat{F}^{+K})(\hat{F}^{-J}\cdot \hat{F}^{-L})\eta_{KL}\eta_{IJ} 
\eea
The first term exactly cancels the second term on substituting the expression for Ricci tensor as given in (\ref{Rmn}). Thus, we finally obtain
\be 
E_c^{(1)} =0.
\ee

\bibliography{references}
\bibliographystyle{jhep}

\end{document}